\documentclass{aa}  

\usepackage{graphicx}
\usepackage{xcolor}
\usepackage[export]{adjustbox}
\usepackage{pdflscape}
\usepackage{txfonts}
\usepackage[pdfpagelabels=false]{hyperref} 
\hypersetup{colorlinks=true,linkcolor=blue,citecolor=blue,filecolor=blue,urlcolor=blue,}
\usepackage{systeme}    
\usepackage{subcaption} 
\usepackage{breqn}      
\usepackage{bm}

\newcommand{\revision}[1]{#1}
\newcommand{\mathrev}[1]{#1}
\begin{document}

   \title{Magnetism in LAMOST CP stars observed by TESS}


   \author{K. Thomson-Paressant\inst{1}\fnmsep\thanks{E-mail: keegan.thomson-paressant@obspm.fr}, C. Neiner\inst{1} \and J. Labadie-Bartz\inst{1}
          }

   \institute{$^1$LESIA, Paris Observatory, PSL University, CNRS, Sorbonne University, Université Paris Cit\'e, 5 place Jules Janssen, 92195  Meudon, France
             }

   \date{Received XXX; accepted XXX}

 
  \abstract
   {
   \revision{A thousand new magnetic candidate CP stars have been identified with LAMOST, among which $\sim$700 prime targets have rotational modulation determined from TESS.}
   }
   {We aim to check for the presence of a magnetic field in a subsample of these LAMOST CP stars, test the viability of the 5200 \AA{} depression used to select the mCP candidates in the LAMOST survey as a reliable indicator of magnetism, and expand on the limited database of known magnetic hot stars. The sample includes \revision{some pulsators that would be valuable targets for magneto-asteroseismology.}
   }
   {We \revision{selected} $\sim$100 magnetic candidate LAMOST CP stars, \revision{presenting} a depression at 5200 \AA{} in their spectrum \revision{and} that also display rotational modulation in their \emph{TESS} photometric lightcurves. We obtained spectropolarimetric observations of 39 targets from this sample with \revision{ESPaDOnS at CFHT}. We utilise the Least Squares Deconvolution method to generate the mean profile of the Stokes V and I parameters, from which the longitudinal magnetic field strength for each target can be determined. 
   \revision{For HD\,49198, we performed} more in-depth analysis to determine the polar magnetic field strength and configuration.
   }
   {We detect fields in at least 36 of our sample of 39 \revision{targets.}
   This success rate in detecting magnetic field (above 92\%) is very high compared to the occurrence of magnetic fields in hot stars ($\sim$10\%). Four of these newly discovered magnetic stars are magnetic pulsators. In particular, we detect the strongest field around a $\delta$ Scuti star discovered to date: a 12 kG dipolar field in HD\,49198.}
   {From our analysis, we conclude that using the 5200 \AA{} depression displayed in the spectra in combination with rotational modulation in photometric data is a very reliable method for identifying magnetic candidates in this population of stars.}


   \maketitle
%

\section{Introduction}
Chemically peculiar (CP) stars are typically main sequence stars that present atypical abundances of certain chemical elements in their surface layers. There are a number of different CP classes, depending on what features they display and what over- or under-abundances they display in their spectra \citep[Ap/Bp, He-strong, He-weak, HgMn, Si, etc.; see][]{preston1974}. It is believed that these chemical peculiarities appear after the formation of the star, as part of its evolution, due to processes such as atomic diffusion or magnetic effects, which dredge chemical elements up to the surface layers or settle in the interior \citep{michaud1970, alecian1986}. CP stars account for approximately 10\% of main sequence intermediate mass stars between early-F and early-B \citep{wolff1968,auriere2007,kochukhov2009,sikora2019}, and a further 10\% of OBA main sequence stars present a strong, large-scale magnetic field \citep{donati2009,ferrario2015,fossati2015}. These two subsets have a significant overlap, corresponding to magnetic CP (mCP) stars with field strengths ranging from $\sim$0.1$-$30 kG \citep{babcock1960,landstreet1982}, mainly of the Ap/Bp-type.

These mCP stars are key observation targets for several reasons. The mechanisms at the origin of their chemical peculiarity (i.e. atomic diffusion, magnetic fields, rotation, and their combinations) are in and of themselves interesting phenomena to study to garner a better understanding of stellar physics and stellar evolution. In addition, their complex atmospheres are great examples upon which we can test and evaluate the validity of atmospheric models, and the variability in their lightcurves due to chemical spots allows for the precise determination of their rotation periods, a benefit that is essential for the study of stellar magnetism \citep{hummerich2018}.

Pulsating stars along the instability strips on the Hertzsprung-Russell (HR) diagram (e.g. $\delta$ Scuti, $\gamma$ Doradus, $\beta$ Cephei, etc.) can also display chemical peculiarities and, since this is typically a good indicator of magnetism, these candidates are very appealing targets for magneto-asteroseismology. Pulsating stars allow us to probe their stellar interiors and physical processes, thus they provide key insight into how stars in general form and evolve throughout the HR diagram \citep{kurtz2022}. Coupling this pulsational information with knowledge of the magnetic field structure for these stars puts strong constraints on seismic models.

Through the magnetic analysis of CP stars presented in this paper, we aim to expand the list of known mCP stars and prove the validity of using the 5200 \AA{} depression for inferring magnetism within this family of stars. This is particularly valuable for the subset of pulsating variable stars contained within this sample, as the number of confirmed magnetic pulsators is very  limited. In addition, a detailed magnetic analysis is performed for one of the target. In-depth analysis of these stars will allow for the constraining of magneto-asteroseismic models, which in turn will provide a better understanding of the influence magnetic fields have on the internal structure of stars and a way to probe the magnetic field inside the star.

\section{Target Selection}
The Large Sky Area Multi-Object Fibre Spectroscopic Telescope \citep[LAMOST,][]{zhao2012,cui2012}, operating at the Xinglong Station in China, was designed to perform a spectroscopic survey of 10 million stars in the Milky Way, as well as millions in other galaxies. Since its first data release in June 2013, it has released additional data sets each year, and has recently surpassed its scientific goals. 

\begin{figure}
    \centering
    \includegraphics[width=\columnwidth]{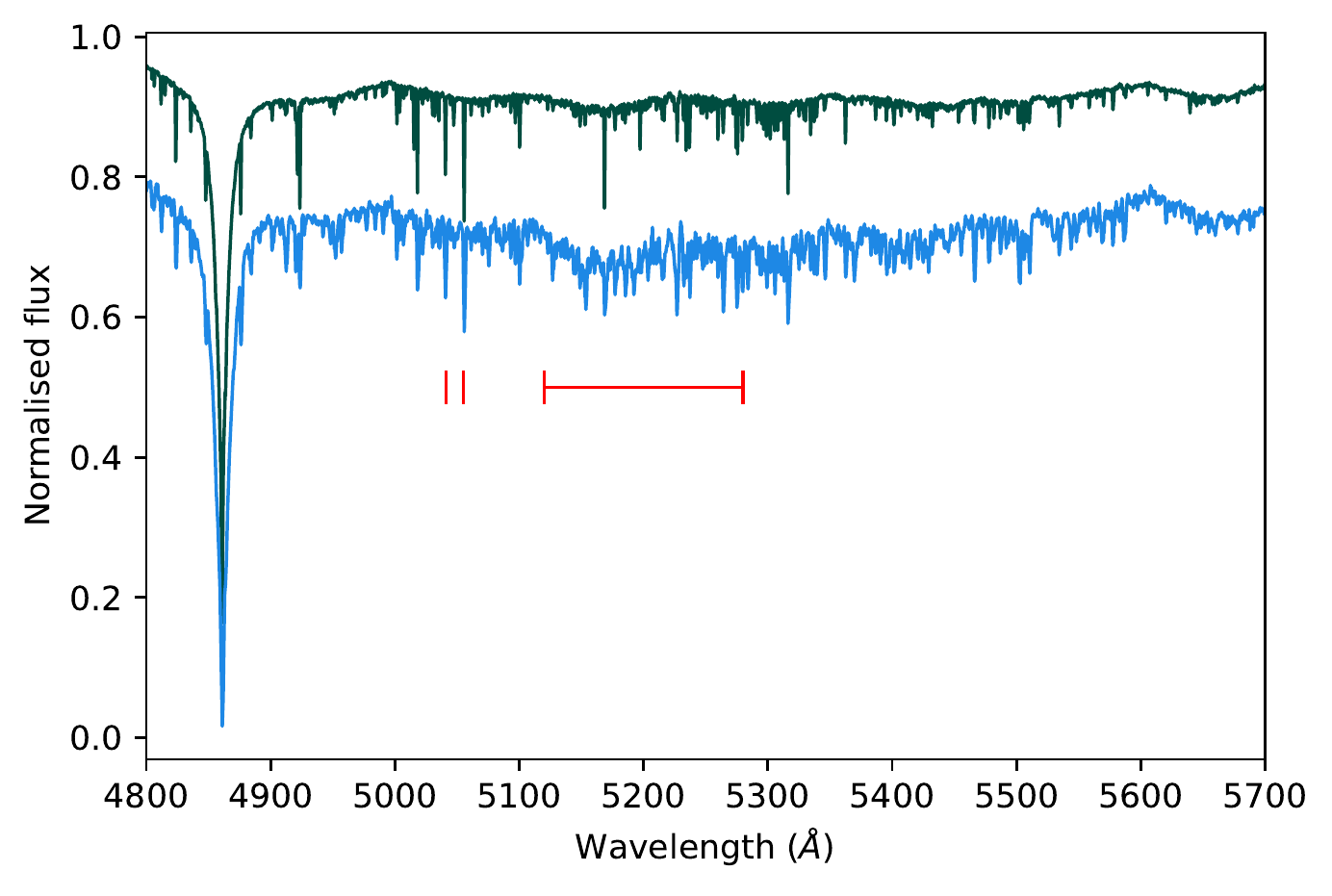}
    \caption{Visualisation of the 5200 \AA{} region (red bar) for non-CP star HD\,220575 (top), and of confirmed mCP star HD\,212714 (bottom). Both stars have a similar spectral type (B8) and their spectra were both taken with the ESPaDOnS instrument. The characteristic Si II lines at 5041 and 5055 \AA{} are also indicated (red tick-marks).}
    \label{fig:depression}
\end{figure}

From the initial LAMOST survey, containing some 4 million stars, a sub-sample of 1002 candidate mCP stars was determined by \citet{hummerich2020}. This was done by selecting all targets that contained a depression at 5200 \AA{} in the LAMOST DR4 spectra of early-type stars (see Fig.~\ref{fig:depression}). This depression at 5200 \AA{}, along with ones at 1400, 1750, 2750, 4100 and 6300 \AA{}, have been determined to be uniquely characteristic of mCP stars \revision{\citep{maitzen1976,kupka2003,kupka2004,khan2007,stigler2014}}. In particular, it was found that Fe is the primary contributor to the 5200 \AA{} depression, and that the latter is present for the whole range of effective temperatures in mCP stars, making it a fantastic indicator of magnetism in these stars. All the stars in this subset are between 100 Myr and 1 Gyr old, with masses between 2-3 $M_{\odot}$.

This subset was then cross-checked with photometry from the Transiting Exoplanet Survey Satellite \citep[\emph{TESS},][]{ricker2015} to look for variability in their lightcurves, and determine stellar parameters such as rotation period and pulsation frequencies \citep{labadie2023}. Through this analysis, they determined the rotation periods for 720 mCP stars, including 25 showing signals consistent with pulsation. 

From here, stars that had $m_V \lesssim 10$, $v \sin i \lesssim 100$ km~s$^{-1}$, required $<$6 hours of observation time with CFHT/ESPaDOnS to achieve a theoretical magnetic field detection threshold of 300 G \emph{and} displayed rotational modulation in their photometric lightcurves were maintained in the selection. From a list of $\sim$100 targets, the sample presented here is finally composed of 39 stars suitable for follow-up spectropolarimetric observations to look for magnetic fields \citep{buysschaert2018,thomson2023}, with a particular emphasis on those found to display pulsation signatures in the previous step. 

\section{Spectropolarimetric Observations}
The targets of our sample were observed with the Echelle SpectroPolarimetric Device for the Observation of Stars \citep[ESPaDOnS,][]{espadons2006}, operating on the Canada France Hawaii Telescope (CFHT) at the Mauna Kea Observatory, Hawaii. The observations primarily took place over 4 semesters between 2021 February 23 and 2022 July 18, but with a few additional observations being performed during January 2024. The list of targets and their respective observations are available in Table~\ref{tab:observations}. The exposure time used for each target was defined by considering the stellar properties (in particular temperature and $v$sin$i$) and assuming that the field possibly present in the targets is at least 300 G. Studies suggest that magnetic OBA stars typically have a magnetic field strength of $\sim$3 kG, though it can range between $\sim$300 G and $\sim$30 kG \citep[e.g.][]{shultz2019}. As a result, the magnetic threshold selected for these observations is expected to be largely sufficient for the overwhelming majority of targets in this sample. In the case of $\delta$\,Sct stars however, recent studies suggest that the average field strength might be closer to $\sim$100 G \citep{thomson2023}, and that of $\gamma$\,Dor might be lower than that, since definite detections of the latter have yet to be discovered.

Data taken by ESPaDOnS was reduced using the \textsc{LibreEsprit} \citep{donati1997} and \textsc{Upena} \citep{martioli2011} pipelines, however we elected to perform continuum normalisation using the \textsc{SpeNT} code \citep{martin2017}. Spectral line masks were computed using data from the Vienna Atomic Line Database \citep[VALD3,][]{piskunov1995,kupka1999,ryabchikova2015} to use as a template, based on the stellar parameters of each respective star, namely $T_{\rm eff}$ and $\log g$. These template masks were fine-tuned by removing regions coinciding with Hydrogen lines, telluric absorption features, and any other features that we considered detrimental to the quality of the spectrum. We then adjusted the line depths of the mask, to better match those found in the stellar spectrum, following the standard procedure described in \citet{grunhut2017}. This is particularly important for CP stars since their line depths differ from those of non-peculiar stars. 

Using the Least Squares Deconvolution method detailed in \citet{donati1997}, we averaged together the spectral lines and Stokes V profiles for each target, resulting in an increased signal-to-noise ratio (SNR) with respect to single lines and improved sensitivity to potential magnetic field signatures. This method works by combining the available lines in each spectrum, and weighting them with respect to their line depth, Landé factor, and corresponding wavelength. LSD also generates an N or `null' profile, which allows to determine whether the signal in V is legitimate or  spurious.

The LSD Stokes profiles for all the targets in our sample are presented in Appendix~\ref{appendix:LSD}.

\begin{table*}[h!]
\centering
\begin{tabular}{ccccccc}
\hline
Target & TIC & Spectral Type & Vmag & Date & Mid-HJD & T$_{\text{exp}}$ \\
 &  &  &  &  & +2459000 & (seqx4xs) \\ \hline
BD\,+00\,2099 & 271310339 & kB8hA3mA3CrEu & 9.90 & 23-Feb-21 & 269.95 & 1x4x519 \\
BD\,+01\,1920 & 271375640 & A1III-IVSrSiEu & 10.02 & 07-Jan-24 & 1318.0421 & 1x4x206 \\
BD\,+08\,2211 & 312111544 & kA3hA5mA7SiEu & 9.81 & 04-Jan-24 & 1315.1807 & 1x4x220 \\
BD\,+10\,2572 & 404536886 & kA3hA7mA9SrCrEu & 9.75 & 21-Feb-21 & 633.0596 & 1x4x133 \\
BD\,+40\,4697 & 305482510 & B9III-IVSi & 9.75 & 18-May-21 & 354.1158 & 1x4x276 \\
BD\,+43\,3648 & 188301298 & kB9hA0mA2Si & 9.57 & 18-May-21 & 354.1043 & 1x4x177 \\
BD\,+44\,767 & 65643991 & kA0hA1mA3(Si) & 9.90 & 25-Nov-21 & 544.7399 & 1x4x73 \\
BD\,+49\,1011 & 428515156 & B8IV-VCrSi & 9.71 & 26-Nov-21 & 545.7688 & 1x4x128 \\
HD\,11140 & 72150546 & B8IVSi & 8.56 & 01-Sept-21 & 459.9184 & 1x4x94 \\
HD\,14251 & 292977419 & B8IV-V & 9.43 & 30-Aug-21 & 457.96 & 1x4x538 \\
HD\,18410 & 251412475 & kB7hA7mA6CrEuSi & 9.14 & 01-Sept-21 & 459.9253 & 2x4x67 \\
HD\,19846 & 445923870 & B9IV-VEu & 8.57 & 25-Nov-21 & 544.7592 & 1x4x359 \\
HD\,22961 & 284084463 & A0IV-VSi & 9.57 & 26-Nov-21 & 545.785 & 1x4x465 \\
HD\,28238 & 373024953 & A2IV-VSrCrEu & 9.16 & 28-Aug-21 & 456.1498 & 1x4x24 \\
 &  &  &  & 26-Nov-21 & 545.8317 & 1x4x24 \\
HD\,36259 & 268068786 & B8IVSi & 9.08 & 01-Sept-21 & 460.0312 & 1x4x878 \\
HD\,36955 & 427377135 & kA0hA2mA4CrEu & 9.58 & 01-Sept-21 & 460.0706 & 1x4x656 \\
HD\,48560 & 11767386 & kA1hA2mA5CrEu & 9.60 & 27-Nov-21 & 547.0746 & 1x4x93 \\
HD\,49198 & 16485771 & A0III-IVCrSi & 9.31 & 26-Nov-21 & 545.8924 & 1x4x319 \\
 &  &  &  & 27-Nov-21 & 547.0635 & 1x4x319 \\
 &  &  &  & 18-Feb-22 & 629.9653 & 1x4x319 \\
 &  &  &  & 21-Feb-22 & 632.8058 & 1x4x319 \\
 &  &  &  & 22-Feb-22 & 633.9703 & 1x4x319 \\
HD\,49522 & 91136550 & A0VCrEuSi & 8.88 & 23-Feb-21 & 269.9116 & 1x4x316 \\
HD\,56514 & 440829763 & A5IV-VSrCrEu & 9.36 & 23-Feb-21 & 269.9287 & 1x4x266 \\
HD\,63843 & 35884762 & A2IVSrSi & 10.25 & 18-Jan-24 & 1329.0059 & 1x4x280 \\
HD\,66533 & 169971995 & kB9hA3mA8SrCrEu & 9.44 & 27-Nov-21 & 547.1427 & 1x4x32 \\
HD\,71047 & 27256691 & A5III-IVSr & 9.60 & 23-Feb-21 & 269.9804 & 1x4x299 \\
HD\,86170 & 62815493 & kA2hA3mA6SrCrEu & 8.41 & 23-Feb-21 & 269.9932 & 1x4x30 \\
HD\,108662 & 393808105 & B9VCrEu & 5.24 & 20-May-21 & 355.9027 & 2x4x30 \\
 &  &  &  & 21-May-21 & 356.8757 & 1x4x30 \\
HD\,177128 & 120495323 & kA1hA4mA6SrCrEu & 9.11 & 18-May-21 & 353.9324 & 1x4x100 \\
HD\,212714 & 164282311 & B8IVEuSi & 8.72 & 18-Jul-22 & 779.8864 & 1x4x1008 \\
HD\,232285 & 240808702 & B9VCrEuSi & 9.42 & 30-Aug-21 & 457.8978 & 1x4x147 \\
HD\,256582 & 319616512 & B5VHeB9 & 10.01 & 22-Feb-22 & 633.8106 & 2x4x1434 \\
HD\,259273 & 234878810 & B9III-IVSi & 9.73 & 25-Dec-21 & 574.8461 & 2x4x937 \\
HD\,266267 & 235391838 & A9VSrSiEu & 10.04 & 15-Jan-24 & 1326.0506 & 1x4x55 \\
HD\,266311 & 237662091 & kA1hA3mA6SrCrEu & 9.74 & 23-Feb-21 & 269.9003 & 1x4x40 \\
HD\,277595 & 122563793 & B8VSi & 9.55 & 01-Sept-21 & 459.9937 & 2x4x200 \\
HD\,281193 & 385555521 & A4IVCrEuSi & 10.08 & 25-Nov-21 & 544.7461 & 1x4x89 \\
TYC\,2873-3205-1 & 384988765 & B9.5VSi & 9.98 & 01-Sept-21 & 459.9566 & 1x4x546 \\
TYC\,3316-892-1 & 458780077 & B9IV-VSi & 9.74 & 25-Nov-21 & 544.7318 & 1x4x186 \\
TYC\,3319-464-1 & 117663254 & B9.5IV-VCrEu & 9.74 & 01-Sept-21 & 459.9384 & 1x4x146 \\
TYC\,3733-133-1 & 252212077 & A2IVCrEu & 9.66 & 26-Nov-21 & 545.8352 & 1x4x40 \\
TYC\,3749-888-1 & 321832920 & A7VSrCrEuSi & 9.81 & 27-Nov-21 & 545.8406 & 1x4x73 \\
\hline
\end{tabular}
\caption{Summary of observational parameters for the targets in our sample. Column 3 provides spectral type from \citet{hummerich2020} and column 4 presents apparent V magnitude. Observation dates are provided in two formats in columns 6 and 7, with number of sequences (seq) and exposure times in seconds in column 8.}
\label{tab:observations}
\end{table*}

\section{Results and discussion}

Using the LSD Stokes V and I profiles generated in the previous step, we calculated the longitudinal field values $B_l$ for each target in our sample \citep{rees1979, wade2000}. To this end, we define a region around the centroid of the line such that it includes the full line profile, while limiting the contribution of the continuum. For those stars for which multiple nights of observation were performed, a $B_l$ value was calculated for each. Table~\ref{tab:results} provides the results of this analysis, with the $B_l$ values appearing in column 8, along with the values for $N_l$ in column 9, calculated from the N profiles by applying the same methods used for calculating $B_l$ from the Stokes V profiles. 

In addition, we applied a False Alarm Probability (FAP) algorithm to each observation, which determines the probability that there is a signal in Stokes V in the same velocity region as used for the $B_l$ calculations above. The FAP algorithm follows the detection criteria described in \citet{donati1992}, such that a definite detection (DD) requires $\text{FAP} \lesssim 10^{-5}$, a marginal detection (MD) corresponds to $10^{-5} \lesssim \text{FAP} \lesssim 10^{-3}$, and a non-detection (ND) is determined to apply for values $\text{FAP} \gtrsim 10^{-3}$. From our sample of 39 stars, 36 of them had definite detections ($>$92\% detection rate), and three had no detections. Of these three, one should not have been included in the initial sample (HD\,14251, see sect.~\ref{sec:hd14251}) and the other two (BD\,+44\,767 and HD\,281193) have lower SNR values than required for a definite detection of a field of at least 300 G, despite large $B_l$ values, and require additional observations to reach our goal threshold and check for the presence of a magnetic field down to 300 G. These 2 stars could thus also be magnetic but our data are of too poor quality to conclude. If we consider only properly selected targets and good quality data, our detection rate is in fact 100\%. 

\revision{Using the determined values for $B_l$, we can then calculate the polar field strength using the following equation adapted  from \citet{schwarzschild1950}:}

\vspace{-10pt}
\begin{equation}
B_{\rm pol} = \frac{4(15-5u)}{15+u} \frac{B_{l}^{\pm}}{\cos(i \mp \beta)}
\label{eq:bpol}
\end{equation}

\revision{where the limb-darkening coefficient $u$ has been determined from \citet{claret2019}, based on the surface gravity and effective temperature of the star in question, and $B_{l}^{+}$/$B_{l}^{-}$ correspond to the maximum/minimum amplitude of a dipolar fit to the $B_l$ measurements, respectively.} 

\revision{With only one or two observations available for most of the targets presented here, only a minimum value of $B_{\text{pol}}$ can be inferred. This is done by assuming $\cos(i \mp \beta) = 1$ and that $B_l^\pm$ is the maximum/minimum amplitude (based on the sign) of the dipolar fit. As such, should additional longitudinal field measurements become available, the resulting $B_{\text{pol}}$ can only increase.}



The $B_{\text{pol}}$ values are presented in Table~\ref{tab:results}. They range from $\sim$300 G to $\sim$13 kG, which is typical for mCP stars \citep[see e.g.][]{shultz2019, sikora2019}.

In order to establish the SNR required to detect magnetic fields with our desired threshold of 300 G, we require values of $v \sin i$ for each target, most of which lacked individual spectroscopic studies to determine accurate values for this parameter. As such, the observations were performed utilising a $v \sin i$ estimate based on the rotational velocity of the star, determined from the precise rotation period from \emph{TESS} photometric lightcurves, while assuming \revision{$\sin i = 1$} and an estimated value for the radius of the star. While this works well as a first approximation, and indeed did not significantly impact the quality of our observations for the vast majority of the targets in our sample, it is not a wholly reliable method. In order to accurately calculate $B_l$ and $B_{\rm pol}$, we elected to determine values for $v \sin i$ ourselves via the Fourier method. This typically operates by performing the Fourier transform of a number well defined and isolated spectral lines, and the position of zeroes present in this Fourier transform provides information on the value of $v \sin i$ \citep{simon-diaz2007}. Thanks to the very high resolution spectra provided by ESPaDOnS and to the nature of the LSD method, we have access to a very well-defined mean spectral line profile for each star from which we can apply this technique, resulting in a $v \sin i$ value that is more representative than performing this step on several individual spectral lines. For the few targets that presented individual studies in the literature, the values we find are consistent with those determined in other works \citep[e.g.][]{wade1997}, excluding large scale surveys such as LAMOST where we found that the determined $v \sin i$ were largely inaccurate and unreliable. The results of our findings are presented in column 5 of Table~\ref{tab:observations}.

\revision{The results of this study provide many new confirmed magnetic stars that are well suited for additional observations and analysis, and also demonstrate the reliability of using the 5200 \AA{} depression in addition to photometric rotational modulation for identifying magnetic candidates.}
Although the number of pulsating stars in this preliminary sample is limited, with three $\delta$\,Sct (HD\,36955, HD\,49198 and HD\,63843) and one SPB (HD\,277595), all of them are confirmed magnetic detections. These magnetic discoveries are significant results since the sample size of known magnetic pulsating stars is small. There are two more pulsators in our sample awaiting observations. As mentioned previously, these magnetic pulsating stars are essential for providing better constraints to magneto-asteroseismological models and consequently improving our understanding of stellar interiors.

A number of special cases, including the few targets with ND results from the FAP algorithm, those determined to contain stellar companions, as well as any stars presenting pulsations, are detailed in the sections below. 

In the few cases where a binary companion was detected and displayed a distinct LSD Stokes I profile from the primary, i.e. SB2 systems, we checked whether a magnetic field could be detected for the secondary. In all such cases, no magnetic field detection was recorded. For these companions, as well as for the 3 target stars with no field detection listed above, we calculate upper limits of the polar magnetic field strength $B_{\rm pol}$, determining the maximum value of a magnetic field that might have remained hidden in the noise of the data. The modelling technique utilised for this calculates 1000 oblique dipole models for each of the LSD Stokes V profiles available for a given target, using random values for inclination $i$, obliquity angle $\beta$, rotational phase, and a white Gaussian noise with a null average and a variance corresponding to the SNR of each profile \citep[for more information, see][]{neiner2015b,thomson2023}. We can then calculate the detection rates from these 1000 models and determine the minimum dipolar magnetic field strength that we had a 90\% probability to detect. This is the detectable upper field limit. 

A couple of stars appear to also contain binary companions, namely HD\,28238 and HD\,49198, determined through the variation in the LSD Stokes I profile in one or multiple observations. These targets are thus identified as SB1 systems. We see additional features and variation in the LSD Stokes I profiles of both HD\,212714 and HD\,256582, however in this case the nature of the variations are less clear, and since the I profiles of the two components are visually indistinct, it makes performing the previously described analysis difficult. We suggest that these features could be the result of rotational effects, or perhaps also SB2 systems.

\subsection{HD\,14251 -- TIC\,292977419}
\label{sec:hd14251}
HD\,14251 was initially flagged as a doubtful CP star during the initial survey of LAMOST targets in \citet{hummerich2020}: they noted that the star shows enhanced metal-lines but no traditional Si, Cr, Sr, or Eu peculiarities. Through the analysis of the spectropolarimetric data and the generation of the Stokes profiles seen in Fig.~\ref{fig:hd14251_lsd}, we clearly see a double I profile characteristic of a spectroscopic binary SB2 system. Neither of the two stars are found to be magnetic, demonstrated by the lack of signal in the Stokes V profile and the ND result of the FAP algorithm. An upper limit calculation gives values of 2250 and 3000 G at 90\% detection probability for the primary and secondary respectively, where for the secondary we used a template mask with an adopted $T_{\rm eff}=8500$ K (compared to 12,500 K for the primary) and determined $v\sin i= 5.44 \pm 2$~km~s$^{-1}$ using the Fourier method centred on the LSD Stokes I profile of the secondary. With these upper field limits, we cannot exclude the possibility that we missed the existence of a magnetic field. However,  our analysis shows that this star is an SB2, which may explain the confusion with a mCP star. It is thus likely that HD\,14251 is simply a non-magnetic SB2 system. 

\subsection{BD+44\,767 -- TIC\,65643991}
Despite a clear and narrow LSD Stokes I profile in Fig.~\ref{fig:bd+44767_lsd}, any signal that could be visible in V is too confounded by noise to be certain, confirmed by the ND result of the FAP algorithm. Upper limit calculations suggest we had a 90\% probability of detecting a field at least as strong as 1500 G, should one exist. Observations were planned using a $v \sin i = 7.86$~km~s$^{-1}$, determined from the \emph{TESS} lightcurves, however using the Fourier method in this work we instead determine a value of $v \sin i = 19\pm 1$~km~s$^{-1}$, which impacts the SNR required to achieve the 300 G magnetic field threshold and would have subsequently required higher exposure times. As a result, the effectiveness of our observations was negatively impacted by an inaccurately determined value of $v \sin i$, further confirmed by the 1500 G upper limit value that we determined. Additional observations with a higher SNR would allow us to confirm or deny the existence of a field in this star. 

\subsection{HD\,281193 -- TIC\,385555521}
This target presented an ND result from the FAP algorithm, in combination with very large error bars in both $B_l$ and $N_l$. There appears to be a possible hint of signal in V in Fig.~\ref{fig:hd281193_lsd}, but it is not statistically significant. Performing upper limit calculations suggests we had a 90\% probability of detecting a field at least as strong as 14,500 G, should one exist, which does not provide much of a restriction. HD\,281193 is a relatively fast rotator with $v \sin i = 103 \pm 3 \; $~km~s$^{-1}$ determined via Fourier analysis. The $v \sin i$ utilised to compute the required SNR for the observations was strongly underestimated which led to the poor quality of our data and high upper limit on the detectable field strength. Unfortunately with such a high $v \sin i$, additional spectropolarimetric observations with enough SNR will be difficult to obtain.

\subsection{HD\,259273 -- TIC\,234878810} \label{sec:HD259273}
This target is a known eclipsing binary (EB) with a 3.409 d orbital period \citep{labadie2023}, however we clearly observe a signal in the Stokes V profile after performing LSD, and acquire a definite detection result from the FAP algorithm. In this case, the magnetic field is clearly aligned on the primary star (as seen in Fig.~\ref{fig:hd259273_lsd}). The N profile also shows a signature at the position of the primary component (FAP gives ND and MD respectively for the two sequences), due to the binary radial velocity motion. It is much weaker than in Stokes V and does not put the magnetic detection into question. However, it may affect the exact $B_l$ value determined from the Stokes V profile. Calculating the FAP for the secondary star gives an ND result, and the upper limits algorithm determined a 90\% probability of detecting a field with $B_{\rm pol}\geq 307$ G. To achieve this, we once again recalculated the LSD profiles using a template mask with $T_{\rm eff} = 8500$ K for the secondary, and determined $v \sin i = 25 \pm 2 \; $~km~s$^{-1}$ from the resulting LSD Stokes I profile using the Fourier method.
For the magnetic primary, we recovered $v \sin i = 5.1 \pm 2 \; $~km~s$^{-1}$, which was considerably slower than anticipated. Additional information for this system is provided in Appendix~\ref{apx:TESS}.

\subsection{HD\,277595 -- TIC\,122563793}
Through our spectropolarimetric analysis, HD\,277595 has also been identified as an SB2 system, once again demonstrated by the double I profile in the LSD Stokes profiles displayed in Fig.~\ref{fig:hd277595_lsd}. In contradiction to the other SB2 HD\,14251, however, HD\,277595 is a clear mCP star in \citet{hummerich2020} and this time we clearly see a signal in Stokes V, in the primary star, implying it is magnetic, confirmed by the definite detection via FAP. We determined a value of $B_l = 526.7 \pm 120.6$ G, and the FAP shows a non-detection for the secondary component of the system, confirming the magnetic field is centred on the primary. Calculating an upper limit value for the secondary, using a $T_{\rm eff} = 8500$ K template mask and calculating $v\sin i = 9.42 \pm 1 $~km~s$^{-1}$ (compared to 12,500 K and $13 \pm 1 \;$~km~s$^{-1}$ for the primary), we determine we had a 90\% probability of detecting a field at least as strong as 1866 G in the secondary component.

In addition, SPB pulsations were observed in the \emph{TESS} lightcurves, which appear to coincide with the magnetic primary, especially considering the B8VSi spectral type and the comparatively cool secondary. The primary component of HD\,277595 is thus a new magnetic SPB star. 

\subsection{HD\,36955 -- TIC\,427377135}
HD\,36955 displays multiple $\delta$\,Sct pulsations between 30 and 50 d$^{-1}$, with an additional peak at 18.5 d$^{-1}$ (see Fig.~\ref{fig:FTs}). Through the analysis of the spectropolarimetric data, we determine a clear magnetic detection, with $B_l = -580.6 \pm 60$ G, corresponding to an approximate polar magnetic field strength of the order of $\mathrev{B_{\rm pol} \gtrsim 1850}$ G, assuming a dipolar field structure. HD\,36955 is thus a new magnetic $\delta$\,Sct star. 

\subsection{HD\,63843 -- TIC\,35884762}
We observe $\delta$\,Sct pulsations in this star with frequencies between 9 and 22 d$^{-1}$ (see Fig.~\ref{fig:FTs}). As with the previous star, we detect a clear magnetic field with $B_l = 2191.3 \pm 15$ G and, assuming dipolar field structure, $\mathrev{B_{\rm pol} \gtrsim 6600}$ G. HD\,63843 is thus another new magnetic $\delta$\,Sct star.

\subsection{HD\,49198 -- TIC\,16485771}
Amongst our sample of stars, one target in particular received more observations than the others: HD\,49198. This was as a result of it being a relatively bright target, displayed clear $\delta$\,Sct pulsations between about 12 and 38 d$^{-1}$ (see Fig.~\ref{fig:FTs}), and had a clear well-defined magnetic detection during the first semester, making it a new magnetic $\delta$\,Sct star and a particularly appealing target for this study. It is also an SB1, as evidenced by the shift in the Stokes I profile seen in Fig.~\ref{fig:hd49198_lsd}. The multiple observations allowed us to perform additional analysis for this star, and infer the overall characterisation of the magnetic field. 

\begin{figure}
    \centering
    \includegraphics[width=0.98\columnwidth]{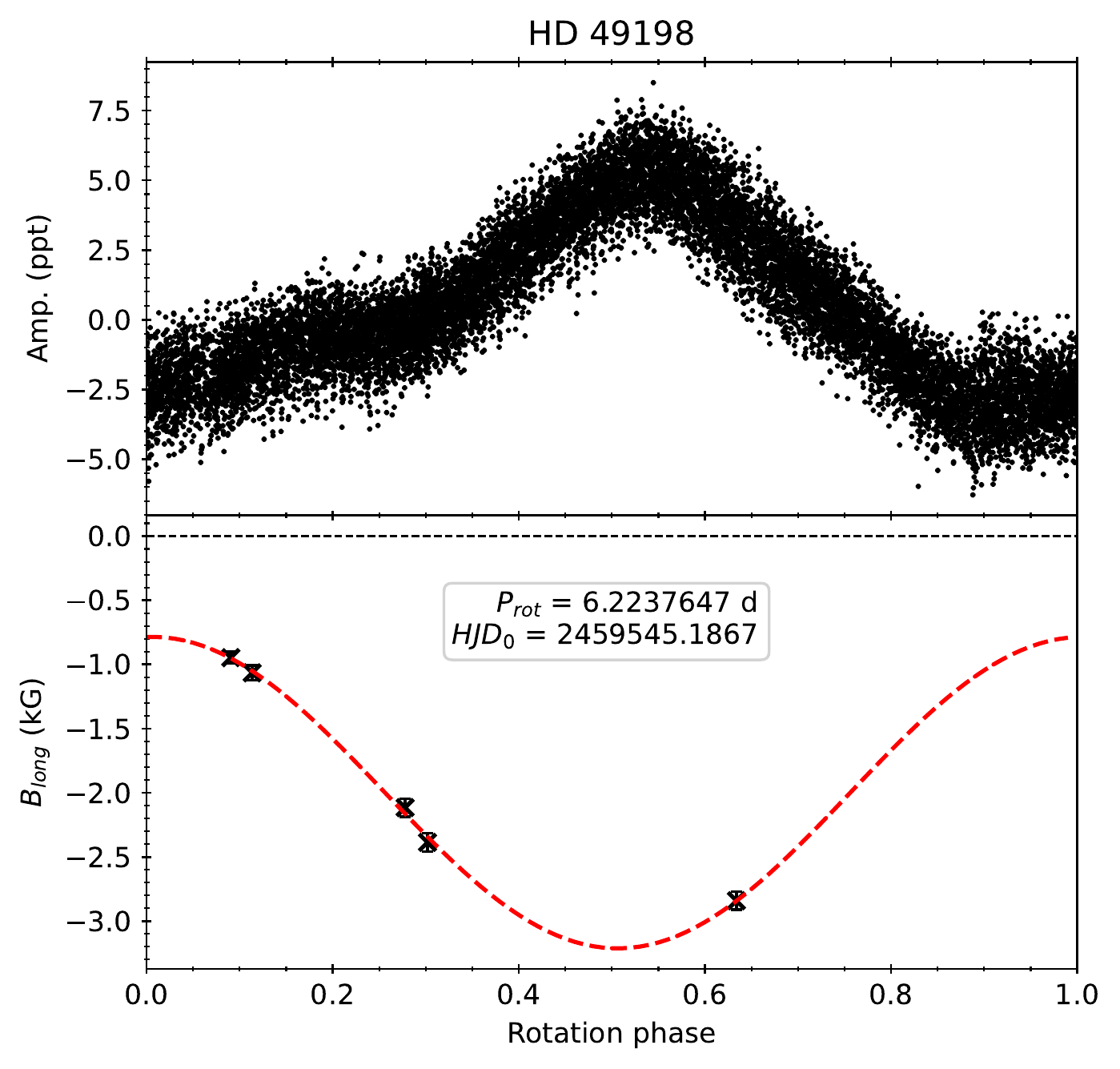}
    \caption{\emph{TESS} photometric lightcurve of HD\,49198 (top panel), phase-folded with respect to the rotation period, utilising data from all currently available sectors (s20, s60, s73). Longitudinal magnetic field values (bottom panel) for HD\,49198 with their respective error bars, phase-folded with respect to the rotation period and overplotted with a simple dipolar fit. The values for the rotation period (P$_{\rm rot}$) and initial Heliocentric Julian Date (${\rm HJD}_0$) are provided for reference.}
    \label{fig:hd49198_bl}
\end{figure}

First, the $B_l$ values we calculated for this star were significantly higher than those for almost any other targets in our sample, with very small error bars. Using these values in combination with the rotational period, inferred by \emph{TESS} photometry, we can visualise the evolution of the longitudinal magnetic field with respect to the rotation of the star, seen in Fig.~\ref{fig:hd49198_bl}. Fitting a simple sinusoidal function to the data points, to approximate the variation of a dipolar field, we see that the fit is extremely representative ($\chi^2 = 0.305$). Such a low $\chi^2$ means we are probably overestimating the error bars on the $B_l$ values. It is therefore safe to assume that the magnetic field of HD\,49198 is dipolar in nature, with a strength varying between:

\[
\left\{\begin{aligned}
B^{+}_{l} &= -785.5 \pm 32.6 \; \text{G} \\ 
B^{-}_{l} &= -3,212.2 \pm 32.6 \; \text{G}
\end{aligned}\right.
\]

Assuming that the field is indeed dipolar, we can derive the polar field strength $B_{\rm pol}$ and the magnetic field obliquity $\beta$. First, we need to calculate the inclination of the rotational axis of the star with respect to the line of sight. For this, we follow the method detailed in various other articles \citep[e.g. ][]{freour2022, thomson2020}, utilising an Oblique Rotator Model first defined by \citet{stibbs1950}, to determine a value for $i$, then for $\beta$ and $B_{\rm pol}$. Using the values from \emph{Gaia} \citep{gaia2023} for the effective temperature $T_{\text{eff}} = 9839 \pm 220\;$K and absolute magnitude $M = 0.72 \pm 0.09$, the value from \emph{TESS} for the rotation period $P_{\rm rot} = 6.2237647 \;$d, inferring the rotational velocity $v \sin i = 14.97 \pm 1\;$~km~s$^{-1}$ using the Fourier method on the LSD Stokes I profile, as well as the radius-luminosity-temperature relation (Eq. 3), we can calculate a value for the inclination:


\vspace{-10pt}
\begin{align}
    \frac{L}{L_\odot} &= 2.51^{(4.83-M)}  \\
    \frac{R}{R_\odot} &= \left(\frac{T}{T_\odot}\right)^{-2} \left(\frac{L}{L_\odot}\right)^{1/2} \\
    i &= \sin^{-1}\left(\frac{v \sin i \, P_{\rm rot}}{2\pi R}\right)
\end{align}

We determine values of $L = 43.92 \pm 3.64 \, L_\odot$, $R = 2.29 \pm 0.14 \, R_\odot$ and $i= 53 \pm 7^\circ$ for the luminosity, radius and inclination respectively. Using this value for $i$ in combination with the ratio between the minimum and maximum longitudinal field values, we can then determine a value for $\beta$ with respect to the inclination \citep{preston1967}. 

\vspace{-10pt}
\begin{equation}
    \frac{B_{l,min}}{B_{l,max}} = \frac{\cos(i+\beta)}{\cos(i-\beta)}
\end{equation}

In our case, utilising the values for $i$, $B^{+}_{l}$, and $B^{-}_{l}$ defined above, we find $\beta = 155 \pm 5^\circ$. Finally, $B_{\rm pol}$ can be estimated utilising Eq.~\ref{eq:bpol} above, taken from \citet{schwarzschild1950}, where we have assumed a limb-darkening coefficient of $u=0.5044$, determined from \citet{claret2019}. Using the calculated values for $i$, $\beta$, and $B_l$, we determine $B_{\rm pol} = 11,740 \pm 2700$ G.
The error on this value is limited by our precision on $i$ and subsequently $\beta$. 

\revision{In Fig.~\ref{fig:hd49198_bl} we see that the maximum of the folded lightcurve (top panel) is slightly shifted with respect to the minimum of the longitudinal field curve (bottom panel). This is often observed in chemically peculiar stars where chemical spots are close to but not right at the magnetic pole. }

This kind of in-depth analysis is most beneficial when working with numerous exposures, as it allows us to better constrain the polar field strength value and its structure, taking pulsations into account. By acquiring additional observations of this star covering the rotation and pulsation phases, we will be able to perform Zeeman Doppler Imaging \citep[ZDI;][]{folsom2017}, allowing us to model and characterise the magnetic field at the surface of the star.

\begin{figure}
    \centering
    \includegraphics[width=0.95\columnwidth]{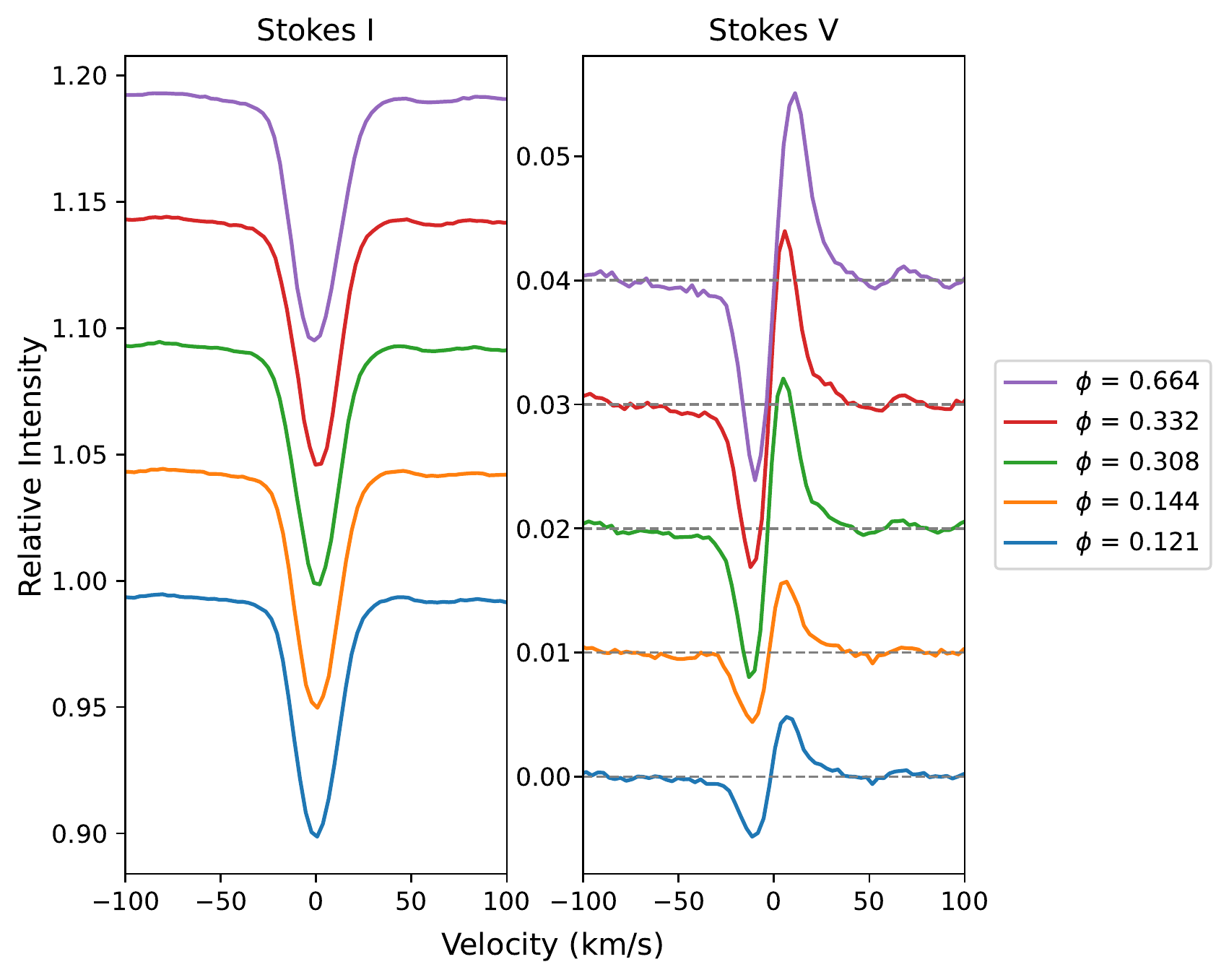}
    \caption{Stokes I (left panel) and V (right panel) profiles for HD\,49198, ordered with respect to the rotational phase of the star and shifted vertically for clarity. The zero-points for each respective Stokes V profile has been defined by a dashed line. }
    \label{fig:hd49198_stokes}
\end{figure}

\section{Conclusions}

Of our sample of 39 candidates we determined 36 of them to be clearly magnetic, with only one likely non-magnetic and a couple requiring better data to check their magnetic status. Thus, a large number of confirmed mCP stars have been determined, suitable for in-depth follow-up and analysis. In particular, three new magnetic $\delta$\,Sct stars and one new magnetic SPB star have been discovered in this work. This is an extremely promising result for the study of stellar magnetism.

This study also proves the viability of the method formalised by \citet{hummerich2020}, using the 5200 \AA{} depression combined with rotational modulation as a reliable indicator for the presence of a magnetic field in CP stars. This is a key result, one that can be utilised to generate large lists of candidate mCP stars with very high chances of detecting fields in all of them.

The magnetic pulsating stars determined as a result of this analysis are essential targets for follow-up observations and subsequent magneto-asteroseismic analysis. Such magnetic pulsating hot stars are rare and our discoveries, while few in number, still provide a large increase of targets in this category. Again, thanks to the reliability of the detection method, all targets that fall in the area overlapped by both chemical peculiarity and pulsation can be quickly and efficiently assessed for the presence of the 5200 \AA{} depression and  rotational modulation and, with a high likelihood, a magnetic field.

\begin{acknowledgements}
We thank S. Hummerich for providing a spreadsheet version of the table in \cite{hummerich2020} before publication of his paper. This work is based on observations obtained at the Canada-France-Hawaii Telescope (CFHT) which is operated by the National Research Council (NRC) of Canada, the Institut National des Sciences de l'Univers of the Centre National de la Recherche Scientifique (CNRS) of France, and the University of Hawaii. This research has made use of the SIMBAD database operated at CDS, Strasbourg (France), and of NASA's Astrophysics Data System (ADS).
This paper includes data collected by the TESS mission, which are publicly available from the Mikulski Archive for Space Telescopes (MAST). Funding for the TESS mission is provided by NASA's Science Mission directorate.
\end{acknowledgements}

\bibliographystyle{aa}
\bibliography{bibliography}

\begin{table*}[htb!]
\centering
\begin{tabular}{@{}c@{\,\,\,}c@{\,\,\,\,}c@{\,\,\,\,}c@{\,\,\,}c@{\,\,\,\,}c@{\,\,\,}c@{\hspace{8pt}}c@{\;}c@{\;}c@{\;}c@{\,}}
\hline
Target & Date & Mn $\lambda$ & Mn & SNR & $v~\sin~i$ & $u$ & FAP & $B_l \pm \sigma B$ & $N_l \pm \sigma N$ & $B_{\rm pol} \pm \sigma B$ or $B_{\rm lim}$ \\
 &  & (nm) & Landé &  & (km~s$^{-1}$) &  &  & (G) & (G) & (G) \\ 
 \hline \hline
BD\,+00\,2099 & 23-Feb-21 & 536.7149 & 1.202 & 1537 & 24 ± 1 & 0.5443 & DD & -521 ± 35 & 34 ± 34 & \revision{ > 1646 ± 111} \\
BD\,+01\,1920 & 07-Jan-24 & 529.6655 & 1.208 & 913 & 5 ± 1 & 0.5443 & DD & -416 ± 7 & -9 ± 6 & \revision{ > 1314 ± 22} \\
BD\,+08\,2211 & 04-Jan-24 & 537.8177 & 1.199 & 943 & 6 ± 1 & 0.5751 & DD & 152 ± 13 & 20 ± 7 & \revision{ > 474 ± 40} \\
BD\,+10\,2572 & 21-Feb-21 & 540.0438 & 1.189 & 1683 & 10 ± 3 & 0.5751 & DD & -1268 ± 26 & 10 ± 20 & \revision{ > 3948 ± 82} \\
BD\,+40\,4697 & 18-May-21 & 535.3467 & 1.197 & 1137 & 18 ± 1 & 0.4444 & DD & 3898 ± 60 & 16 ± 37 & \revision{ > 12,900 ± 200} \\
BD\,+43\,3648 & 18-May-21 & 531.1814 & 1.205 & 1520 & 17 ± 1 & 0.5046 & DD & -2021 ± 106 & 6 ± 92 & \revision{ > 6504 ± 341} \\
BD\,+44\,767 & 25-Nov-21 & 555.8105 & 1.199 & 1213 & 18 ± 1 & 0.5151 & ND & 453 ± 192 & -40 ± 193 & < 1500 \\
BD\,+49\,1011 & 26-Nov-21 & 543.1386 & 1.192 & 801 & 7 ± 1 & 0.4249 & DD & -279 ± 38 & -35 ± 37 & \revision{ > 934 ± 128} \\
HD\,11140 & 01-Sept-21 & 533.8171 & 1.191 & 2174 & 21 ± 1 & 0.4249 & DD & 206 ± 70 & 82 ± 70 & \revision{ > 690 ± 234} \\
HD\,14251a & 30-Aug-21 & 532.7064 & 1.191 & 3416 & 40 ± 1 & 0.4249 & ND & 142 ± 152 & 82 ± 152 & < 2250 \\
HD\,14251b &  &  &  &  & $5 \pm 2$ &  & ND & 2 ± 89 & -35 ± 89 & < 3000 \\
HD\,18410 & 01-Sept-21 & 542.0123 & 1.192 & 1478 & 21 ± 1 & 0.5751 & DD & 1867 ± 39 & -7 ± 34 & \revision{ > 5814 ± 121} \\
HD\,19846 & 25-Nov-21 & 528.7895 & 1.197 & 1360 & 27 ± 1 & 0.4444 & DD & 194 ± 50 & -48 ± 35 & \revision{ > 642 ± 165} \\
HD\,22961 & 26-Nov-21 & 529.6589 & 1.205 & 1839 & 18 ± 1 & 0.5046 & DD & 102 ± 38 & 45 ± 38 & \revision{ > 328 ± 124} \\
HD\,28238 & 28-Aug-21 & 537.5226 & 1.206 & 747 & 6 ± 1 & 0.5279 & DD & -207 ± 17 & 0 ± 16 & \revision{ > 684 ± 55} \\
 & 26-Nov-21 & 536.1829 & 1.207 & 822 &  &  & DD & 1 ± 18 & 17 ± 18 &  \\
HD\,36259 & 01-Sept-21 & 533.7634 & 1.191 & 2190 & 28 ± 1 & 0.4249 & DD & 761 ± 19 & -1 ± 17 & \revision{ > 2548 ± 63} \\
HD\,36955 & 01-Sept-21 & 526.764 & 1.208 & 1855 & 33 ± 1 & 0.5279 & DD & -581 ± 60 & 35 ± 58 & \revision{ > 1849 ± 191} \\
HD\,48560 & 27-Nov-21 & 529.8405 & 1.207 & 1466 & 8 ± 2 & 0.5279 & DD & -932 ± 30 & 14 ± 28 &>  2967 ± 94 \\
HD\,49198 & 26-Nov-21 & 525.9691 & 1.206 & 1459 & 15 ± 1 & 0.5044 & DD & -1065 ± 59 & -56 ± 50 & 11,740 ± 2700 \\
 & 27-Nov-21 & 524.039 & 1.206 & 1394 &  &  & DD & -2386 ± 72 & -26 ± 41 &  \\
 & 18-Feb-22 & 527.567 & 1.205 & 1220 &  &  & DD & -2841 ± 70 & -50 ± 38 &  \\
 & 21-Feb-22 & 523.3311 & 1.206 & 1454 &  &  & DD & -947 ± 44 & 41 ± 35 &  \\
 & 22-Feb-22 & 528.355 & 1.205 & 1392 &  &  & DD & -2115 ± 70 & 40 ± 45 &  \\
HD\,49522 & 23-Feb-21 & 526.4611 & 1.205 & 1221 & 22 ± 1 & 0.5046 & DD & -2353 ± 25 & -7 ± 16 & \revision{ > 7573 ± 80} \\
HD\,56514 & 23-Feb-21 & 531.4868 & 1.216 & 2027 & 32 ± 1 & 0.5765 & DD & -256 ± 54 & 1 ± 53 & \revision{ > 795 ± 168} \\
HD\,63843 & 18-Jan-24 & 526.345 & 1.189 & 1708 & 7 ± 2 & 0.5659 & DD & 2191 ± 15 & 5 ± 10 & \revision{ > 6853 ± 47} \\
HD\,66533 & 27-Nov-21 & 532.1027 & 1.197 & 906 & 15 ± 1 & 0.5443 & DD & -282 ± 19 & -17 ± 18 & \revision{ > 889 ± 59} \\
HD\,71047 & 23-Feb-21 & 525.6292 & 1.199 & 1816 & 16 ± 1 & 0.5765 & DD & 165 ± 28 & -1 ± 25 & \revision{ > 516 ± 88} \\
HD\,86170 & 23-Feb-21 & 534.0372 & 1.197 & 924 & 5 ± 2 & 0.5443 & DD & -104 ± 26 & -16 ± 24 & \revision{ > 328 ± 83} \\
HD\,108662 & 20-May-21 & 534.5184 & 1.195 & 1394 & 20 ± 1 & 0.4444 & DD & -690 ± 73 & 8 ± 73 & \revision{ > 2283 ± 243} \\
 & 21-May-21 & 527.9288 & 1.197 & 1118 &  &  & DD & -5 ± 22 & -32 ± 20 &  \\
HD\,177128 & 18-May-21 & 527.0712 & 1.200 & 1452 & 19 ± 1 & 0.5630 & DD & 860 ± 58 & -67 ± 54 & \revision{ > 2693 ± 183} \\
HD\,212714 & 18-Jul-22 & 543.2427 & 1.181 & 2961 & 59 ± 7 & 0.4249 & DD & -114 ± 38 & -8 ± 38 & \revision{ > 380 ± 128} \\
HD\,232285 & 30-Aug-21 & 534.9881 & 1.197 & 925 & 7 ± 1 & 0.4444 & DD & 771 ± 17 & 12 ± 13 & \revision{ > 2553 ± 56} \\
HD\,256582 & 22-Feb-22 & 566.5622 & 1.197 & 3907 & 45 ± 1 & 0.3749 & DD & -154 ± 74 & -24 ± 73 & \revision{ > 526 ± 253} \\
HD\,259273a & 25-Dec-21 & 528.9497 & 1.197 & 2452 & 5 ± 2 & 0.4444 & DD & -144 ± 8 & 0 ± 8 & \revision{ > 449 ± 26} \\
HD\,259273b &  &  &  &  & 25 ± 2 &  & ND & -2000 ± 990 & 538 ± 824 & < 307 \\
HD\,266267 & 15-Jan-24 & 540.7631 & 1.208 & 919 & 5 ± 3 & 0.5765 & DD & 403 ± 11 & 27 ± 11 & \revision{ > 1254 ± 35} \\
HD\,266311 & 23-Feb-21 & 539.0441 & 1.196 & 767 & 8 ± 1 & 0.5443 & DD & -1347 ± 21 & -32 ± 18 & \revision{ > 4256 ± 65} \\
HD\,277595a & 01-Sept-21 & 539.4005 & 1.191 & 3635 & 13 ± 1 & 0.4249 & DD & 527 ± 121 & 13 ± 120 & \revision{ > 1764 ± 404} \\
HD\,277595b &  &  &  &  & 9 ± 1 &  & ND & 20 ± 121 & -7 ± 121 & < 1866 \\
HD\,281193 & 25-Nov-21 & 537.4729 & 1.206 & 2296 & 104 ± 2 & 0.5630 & ND & 628 ± 321 & -638 ± 322 & < 14,500 \\
TYC\,2873-3205-1 & 01-Sept-21 & 526.6307 & 1.201 & 1815 & 18 ± 1 & 0.4724 & DD & 213 ± 35 & -1 ± 35 & \revision{ > 697 ± 114} \\
TYC\,3316-892-1 & 25-Nov-21 & 538.0924 & 1.197 & 1648 & 18 ± 1 & 0.4444 & DD & 965 ± 57 & -55 ± 53 & \revision{ > 3192 ± 187} \\
TYC\,3319-464-1 & 01-Sept-21 & 542.2375 & 1.198 & 614 & 7 ± 1 & 0.4444 & DD & 280 ± 22 & 13 ± 19 & \revision{ > 925 ± 72} \\
TYC\,3733-133-1 & 26-Nov-21 & 548.7484 & 1.197 & 754 & 9 ± 1 & 0.5279 & DD & -1127 ± 30 & -13 ± 24 & \revision{ > 3589 ± 96} \\
TYC\,3749-888-1 & 27-Nov-21 & 555.074 & 1.188 & 2043 & 25 ± 1 & 0.5751 & DD & -458 ± 63 & -47 ± 62 & \revision{ > 1425 ± 196} \\
\hline
\end{tabular}
\caption{Results of magnetic characterisation of our sample. Columns 3, 4, and 5 display the mean values for wavelength and Landé factor, and SNR for the LSD I profiles respectively. In the case of multiple sequences during a single night, these values were averaged together. Our Fourier-method determined values for $v~\sin~i$ are presented in column 6, along with in column 7 the determined limb-darkening coefficients $u$ \citep{claret2019} for $B_{\rm pol}$ calculations. The results of the FAP algorithm are shown in column 8, and the longitudinal field measurements from the Stokes V ($B_l$) and the corresponding N ($N_l$) profiles, with their respective errors, are provided in columns 9 and 10. Finally, values for the polar field strength $B_{\rm pol}$ and their errors when a magnetic field was detected or the upper limit $B_{\rm lim}$ for non-detections are presented in column 11. \revision{The $B_{\text pol}$ values for targets having only one or two polarimetric sequences are lower bounds, and have been represented as such.}}
\label{tab:results}
\end{table*}

\clearpage
\begin{appendix}
{\onecolumn
\section{LSD profiles}
\label{appendix:LSD}

In this appendix we present the LSD profiles of each of the 39 targets. 

\begin{figure}[!ht]
    \centering
    \begin{subfigure}{0.49\textwidth}
         \centering
         \includegraphics[width=\textwidth]{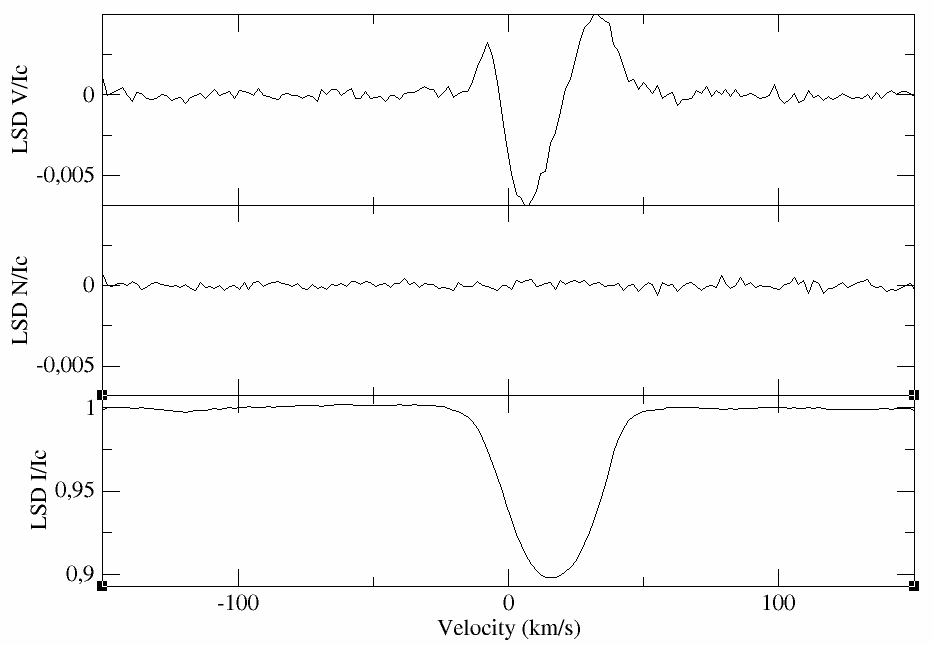}
         \caption{BD+00 2099 -- TIC 271310339}
         \label{fig:bd+002099_lsd}
     \end{subfigure}
     \hfill
     \begin{subfigure}{0.49\textwidth}
         \centering
         \includegraphics[width=\textwidth]{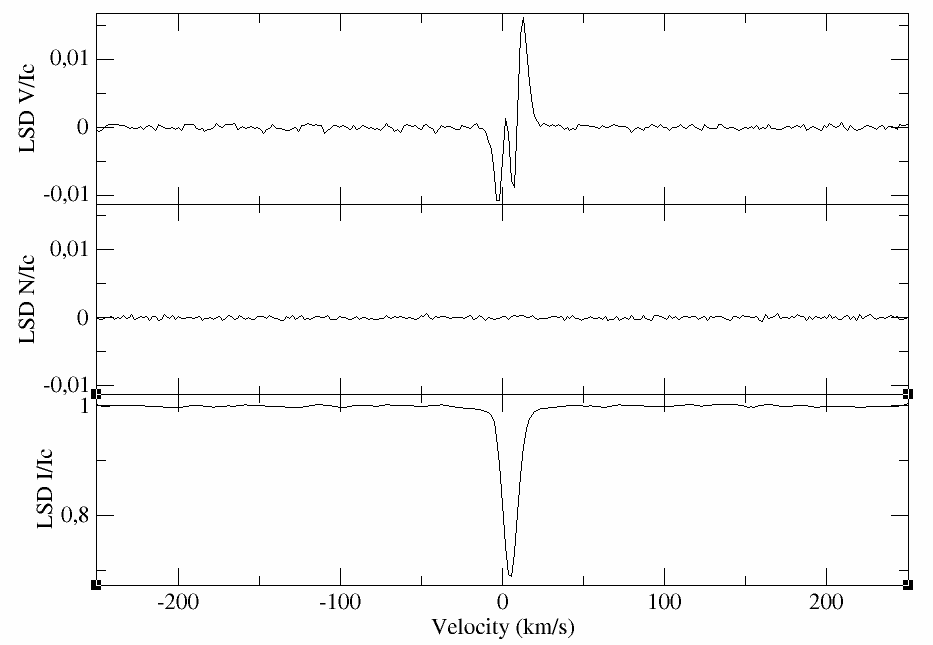}
         \caption{BD+01 1920 -- TIC 271375640}
         \label{fig:bd+011920_lsd}
     \end{subfigure}
     \vskip\baselineskip
     \begin{subfigure}{0.49\textwidth}
         \centering
         \includegraphics[width=\textwidth]{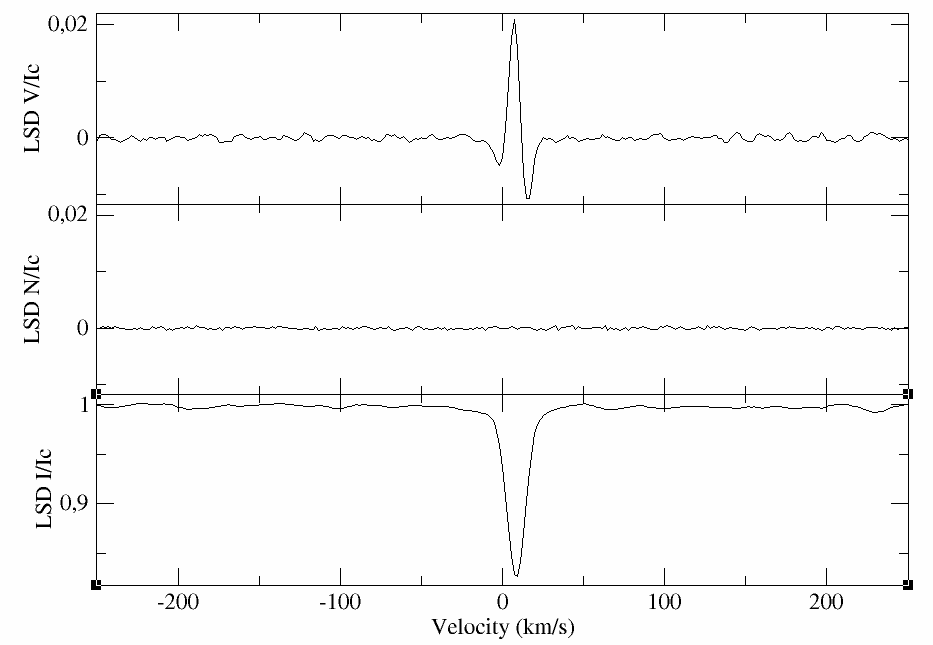}
         \caption{BD+08 2211 -- TIC 312111544}
         \label{fig:bd+082211_lsd}
     \end{subfigure}
     \hfill
     \begin{subfigure}{0.49\textwidth}
         \centering
         \includegraphics[width=\textwidth]{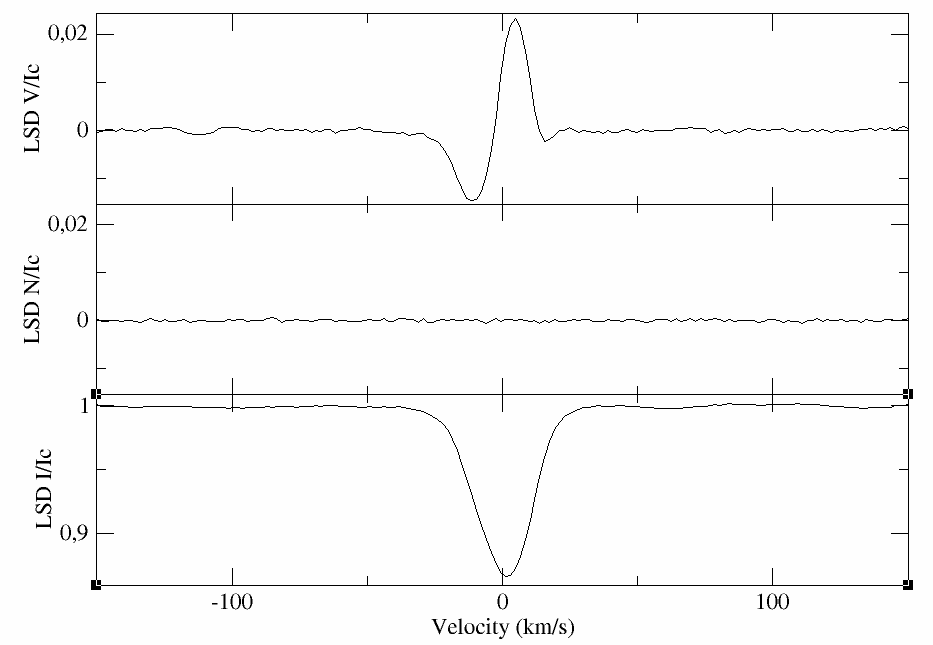}
         \caption{BD+10 2572 -- TIC 404536886}
         \label{fig:bd+102572_lsd}
     \end{subfigure}
     \vskip\baselineskip
     \begin{subfigure}{0.49\textwidth}
         \centering
         \includegraphics[width=\textwidth]{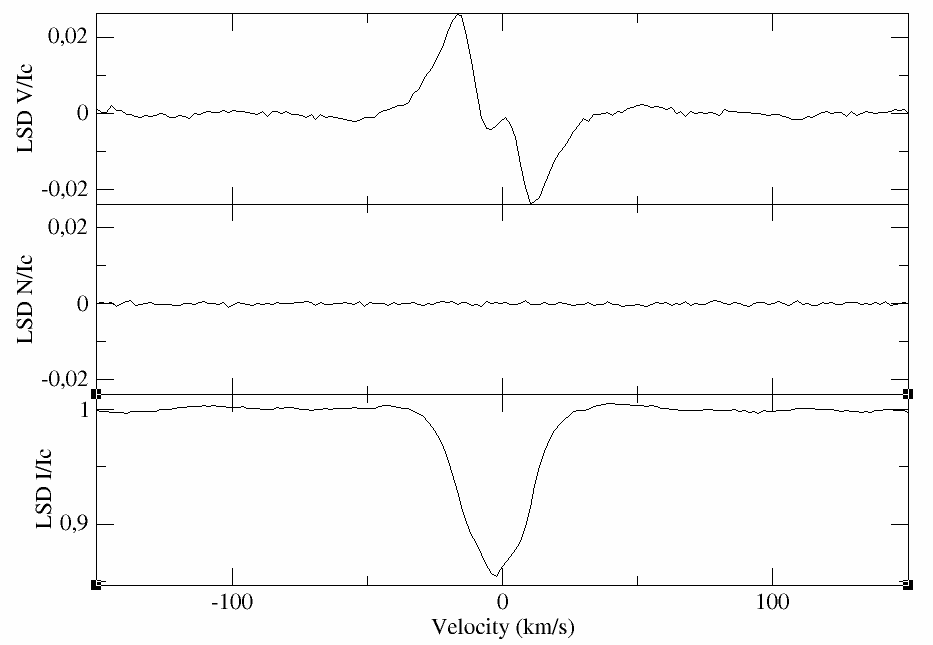}
         \caption{BD+40 4697 -- TIC 305482510}
         \label{fig:bd+404697_lsd}
     \end{subfigure}
     \hfill
     \begin{subfigure}{0.49\textwidth}
         \centering
         \includegraphics[width=\textwidth]{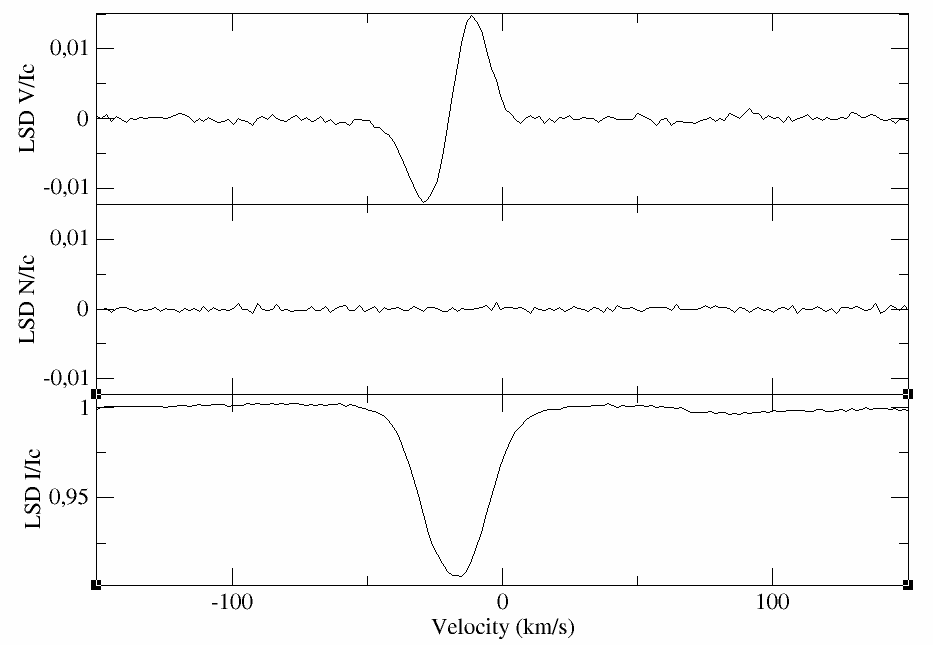}
         \caption{BD+43 3648 -- TIC 188301298}
         \label{fig:bd+433684_lsd}
     \end{subfigure}
    \caption{LSD Stokes V (upper panel), N (middle panel), and Stokes I (bottom panel) profiles for the various stars in our sample, arranged by stellar identifier.}
    \label{fig:stokes1}
\end{figure}

\begin{figure}[!ht]
    \centering
    \begin{subfigure}{0.49\textwidth}
         \centering
         \includegraphics[width=\textwidth]{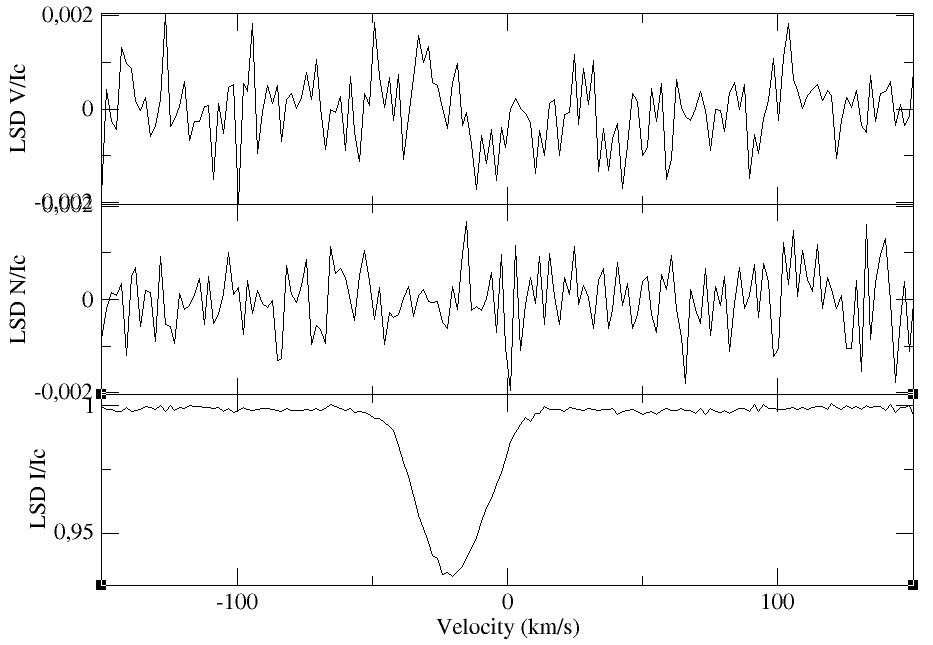}
         \caption{BD+44 767 -- TIC 65643991}
         \label{fig:bd+44767_lsd}
     \end{subfigure}
     \hfill
     \begin{subfigure}{0.49\textwidth}
         \centering
         \includegraphics[width=\textwidth]{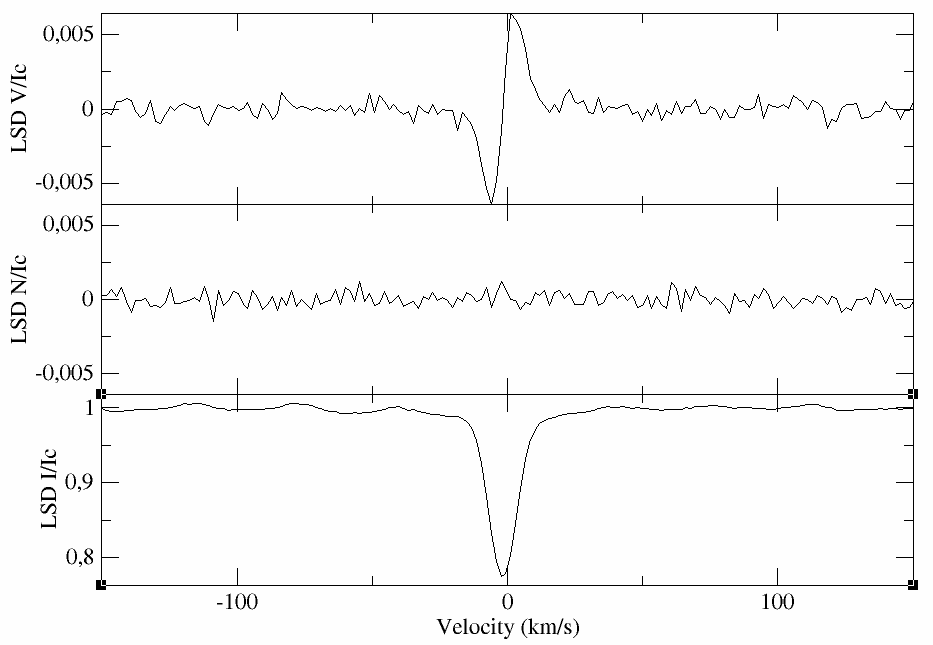}
         \caption{BD+49 1011 -- TIC 428515156}
         \label{fig:bd+491011_lsd}
     \end{subfigure}
     \vskip\baselineskip
     \begin{subfigure}{0.49\textwidth}
         \centering
         \includegraphics[width=\textwidth]{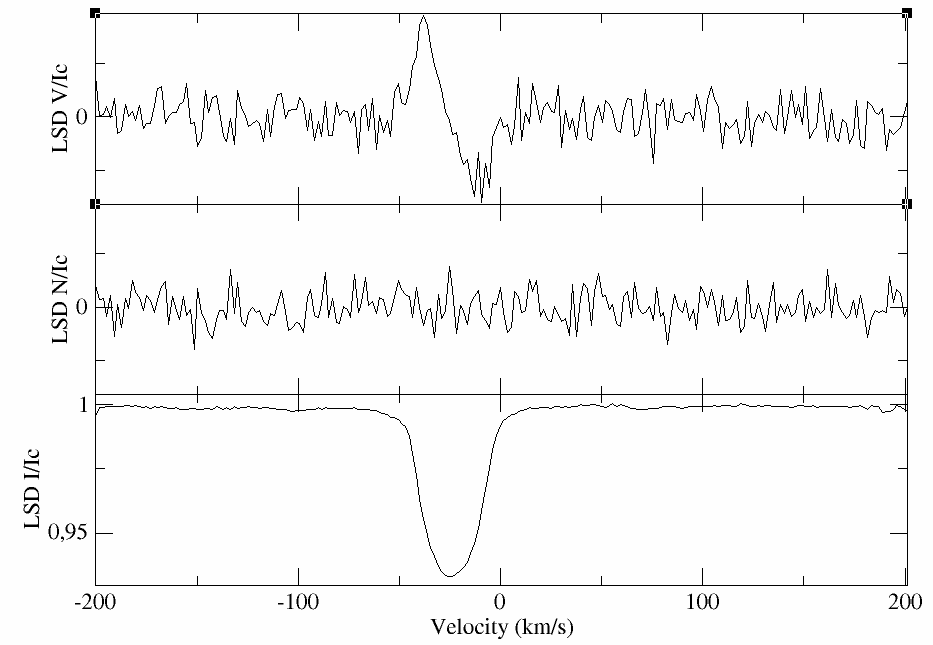}
         \caption{HD\,11140 -- TIC 72150546}
         \label{fig:hd11140_lsd}
     \end{subfigure}
     \hfill
     \begin{subfigure}{0.49\textwidth}
         \centering
         \includegraphics[width=\textwidth]{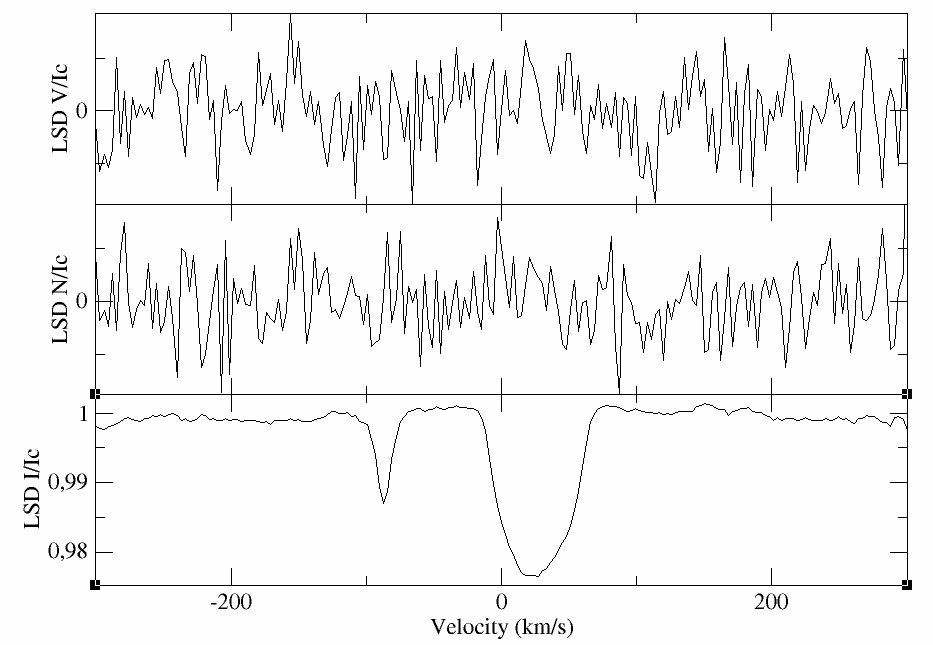}
         \caption{HD\,14251 -- TIC 292977419}
         \label{fig:hd14251_lsd}
     \end{subfigure}
     \vskip\baselineskip
     \begin{subfigure}{0.49\textwidth}
         \centering
         \includegraphics[width=\textwidth]{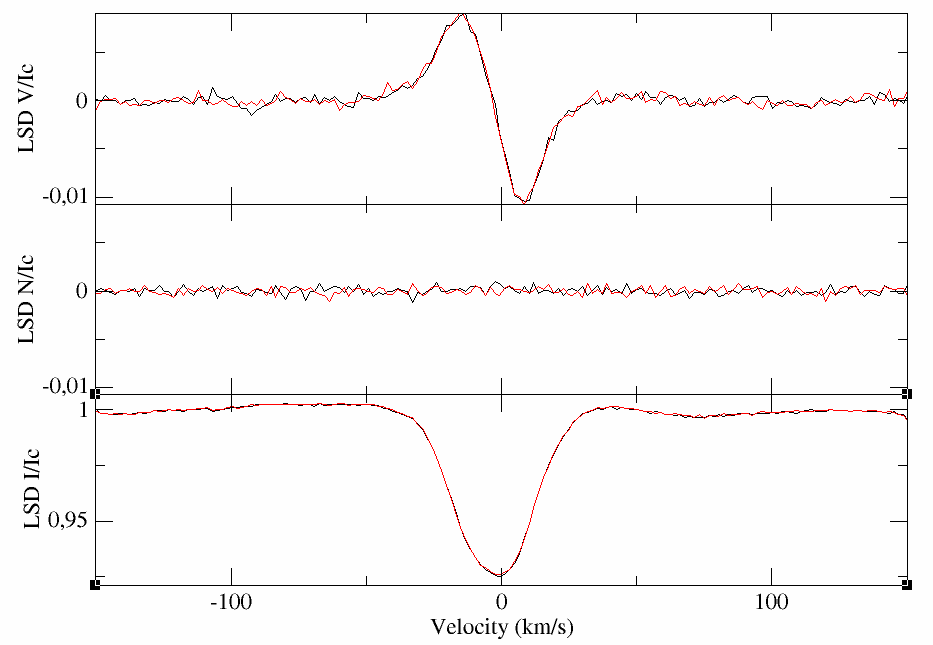}
         \caption{HD\,18410 -- TIC 251412475}
         \label{fig:hd18410_lsd}
     \end{subfigure}
     \hfill
     \begin{subfigure}{0.49\textwidth}
         \centering
         \includegraphics[width=\textwidth]{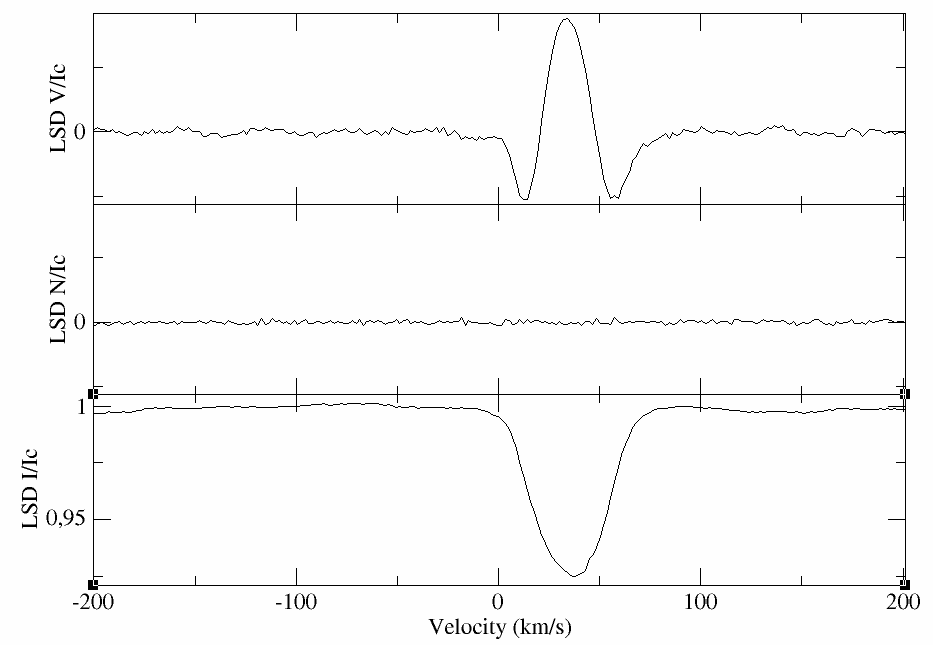}
         \caption{HD\,19846 -- TIC 445923870}
         \label{fig:hd19846_lsd}
     \end{subfigure}
    \caption{LSD Stokes V (upper panel), N (middle panel), and Stokes I (bottom panel) profiles for the various stars in our sample, arranged by stellar identifier (cont.).}
    \label{fig:stokes2}
\end{figure}

\begin{figure}[!ht]
    \centering
    \begin{subfigure}{0.49\textwidth}
         \centering
         \includegraphics[width=\textwidth]{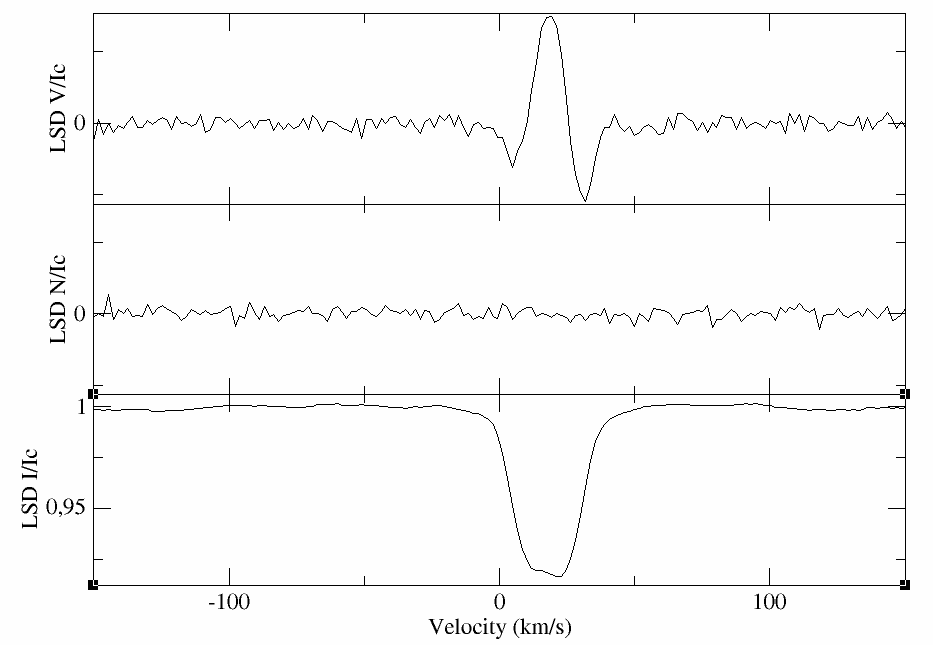}
         \caption{HD\,22961 -- TIC 284084463}
         \label{fig:hd22961_lsd}
     \end{subfigure}
     \hfill
     \begin{subfigure}{0.49\textwidth}
         \centering
         \includegraphics[width=\textwidth]{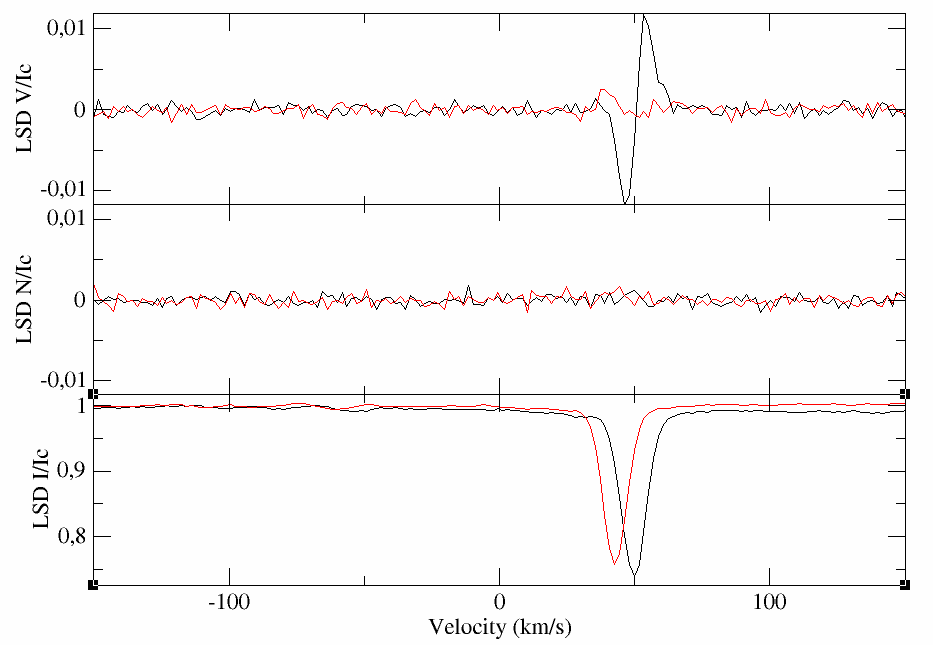}
         \caption{HD\,28238 -- TIC 373024953}
         \label{fig:hd28238_lsd}
     \end{subfigure}
     \vskip\baselineskip
     \begin{subfigure}{0.49\textwidth}
         \centering
         \includegraphics[width=\textwidth]{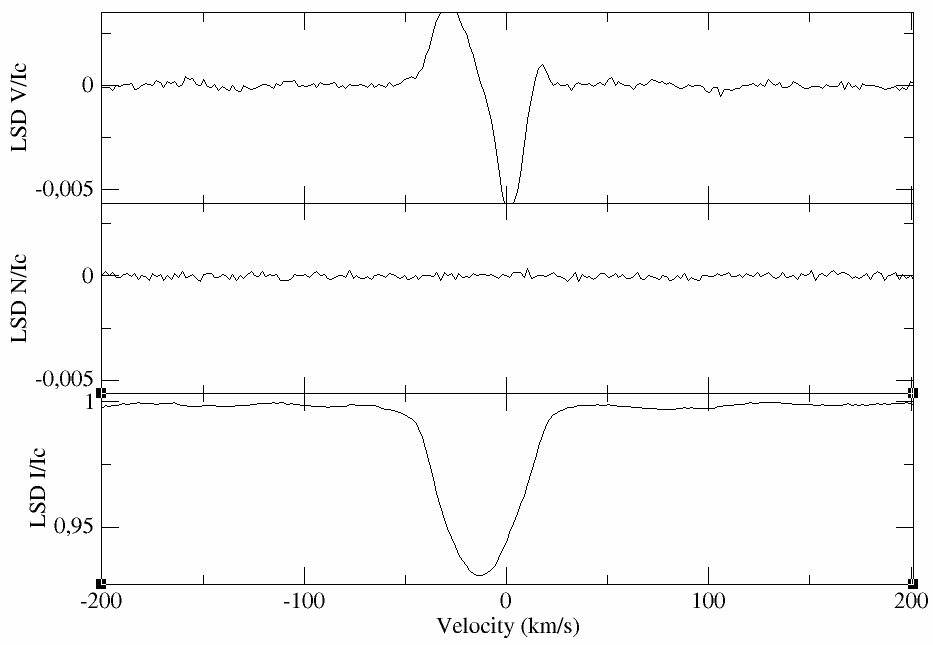}
         \caption{HD\,36259 -- TIC 268068786}
         \label{fig:hd36259_lsd}
     \end{subfigure}
     \hfill
     \begin{subfigure}{0.49\textwidth}
         \centering
         \includegraphics[width=\textwidth]{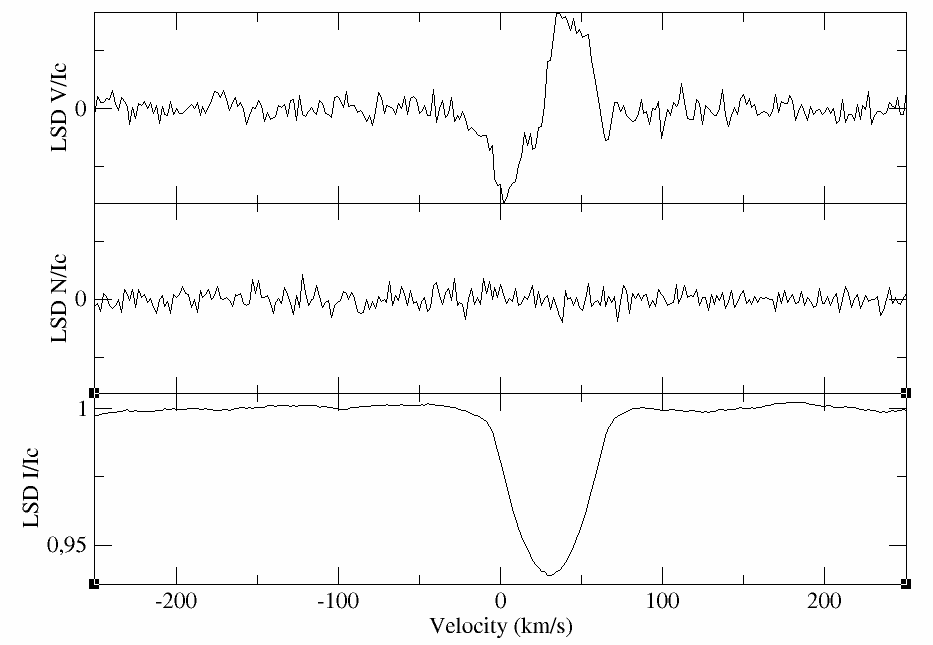}
         \caption{HD\,36955 -- TIC 427377135}
         \label{fig:hd36955_lsd}
     \end{subfigure}
     \vskip\baselineskip
     \begin{subfigure}{0.49\textwidth}
         \centering
         \includegraphics[width=\textwidth]{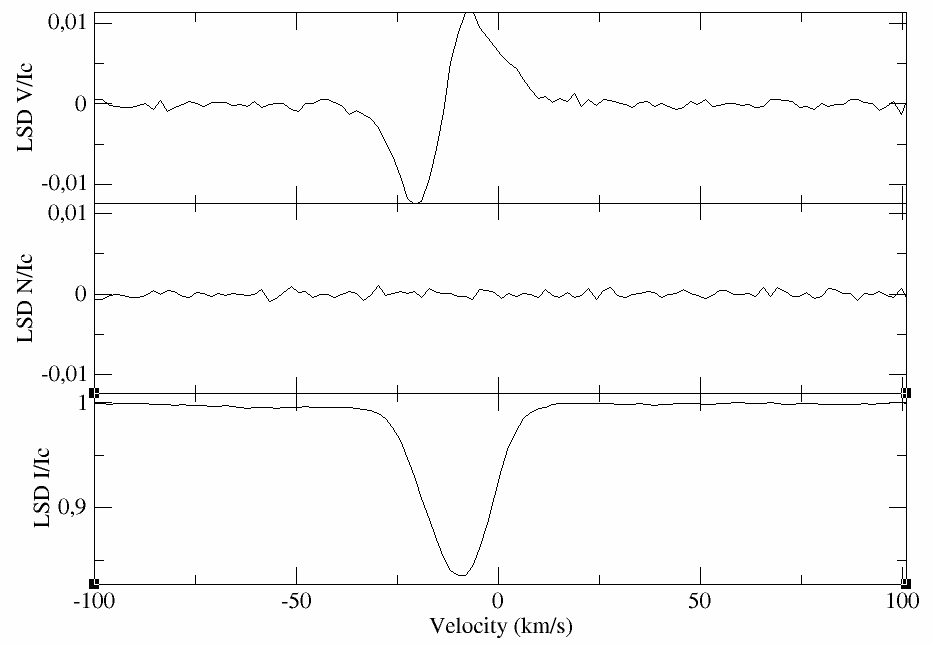}
         \caption{HD\,48560 -- TIC 11767386}
         \label{fig:hd48560_lsd}
     \end{subfigure}
     \hfill
     \begin{subfigure}{0.49\textwidth}
         \centering
         \includegraphics[width=\textwidth]{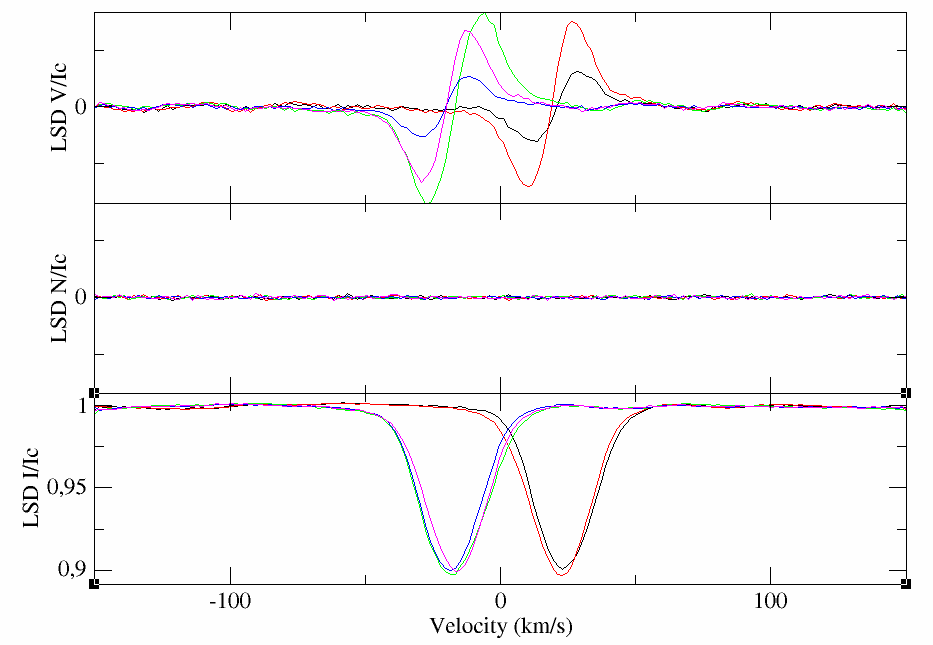}
         \caption{HD\,49198 -- TIC 16485771}
         \label{fig:hd49198_lsd}
     \end{subfigure}
    \caption{LSD Stokes V (upper panel), N (middle panel), and Stokes I (bottom panel) profiles for the various stars in our sample, arranged by stellar identifier (cont.).}
    \label{fig:stokes3}
\end{figure}

\begin{figure}[!ht]
    \centering
    \begin{subfigure}{0.49\textwidth}
         \centering
         \includegraphics[width=\textwidth]{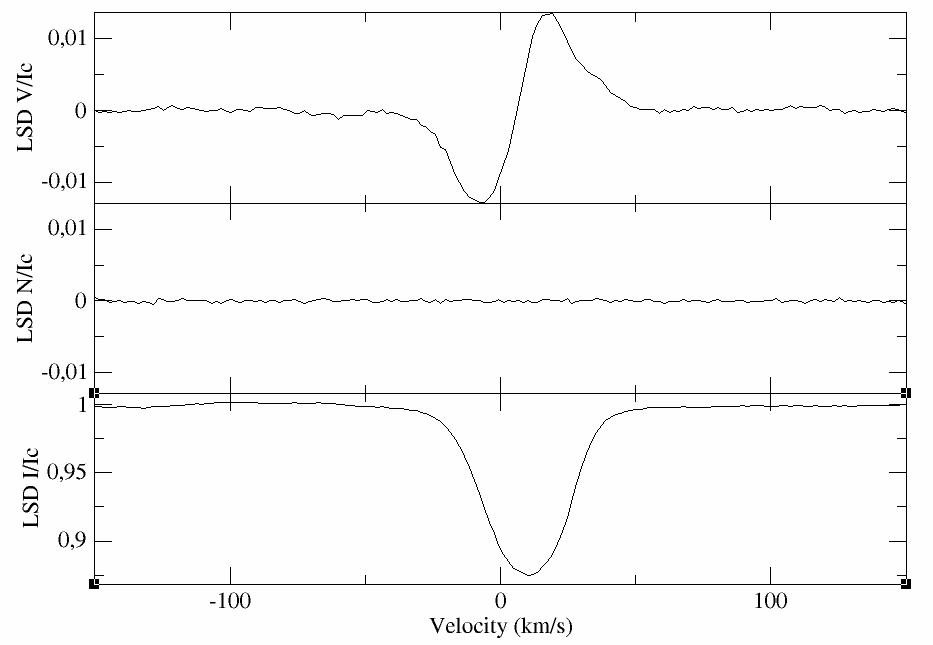}
         \caption{HD\,49522 -- TIC 91136550}
         \label{fig:hd49522_lsd}
     \end{subfigure}
     \hfill
     \begin{subfigure}{0.49\textwidth}
         \centering
         \includegraphics[width=\textwidth]{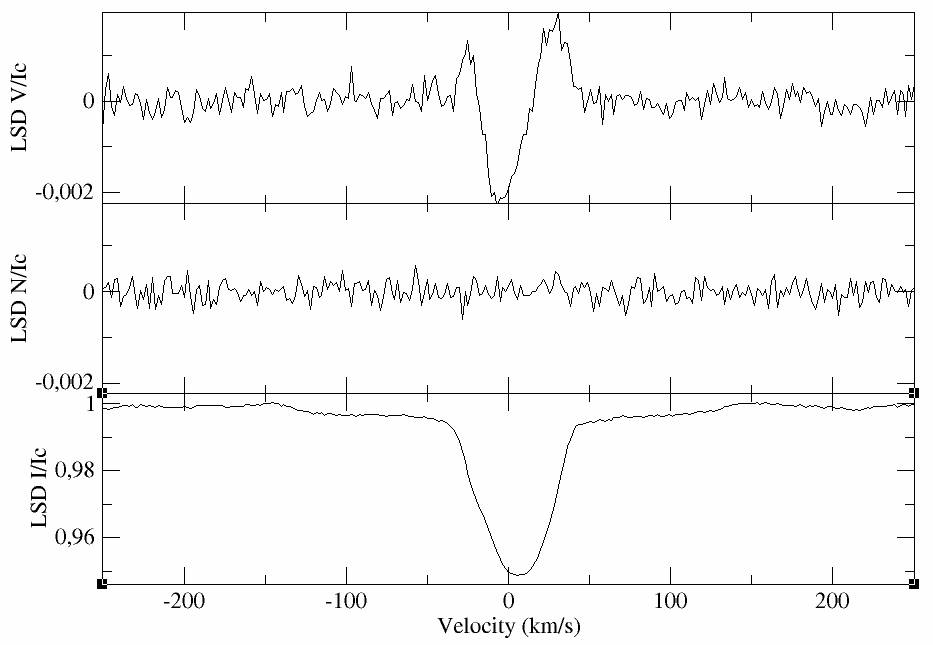}
         \caption{HD\,56514 -- TIC 440829763}
         \label{fig:hd56514_lsd}
     \end{subfigure}
     \vskip\baselineskip
     \begin{subfigure}{0.49\textwidth}
         \centering
         \includegraphics[width=\textwidth]{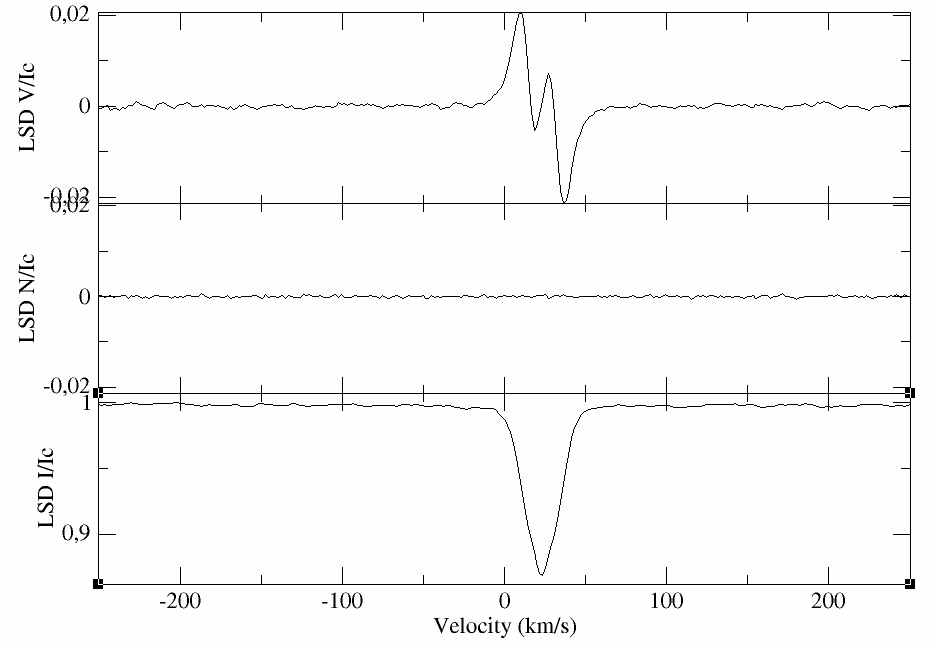}
         \caption{HD\,63843 -- TIC 35884762}
         \label{fig:hd63843_lsd}
     \end{subfigure}
     \hfill
     \begin{subfigure}{0.49\textwidth}
         \centering
         \includegraphics[width=\textwidth]{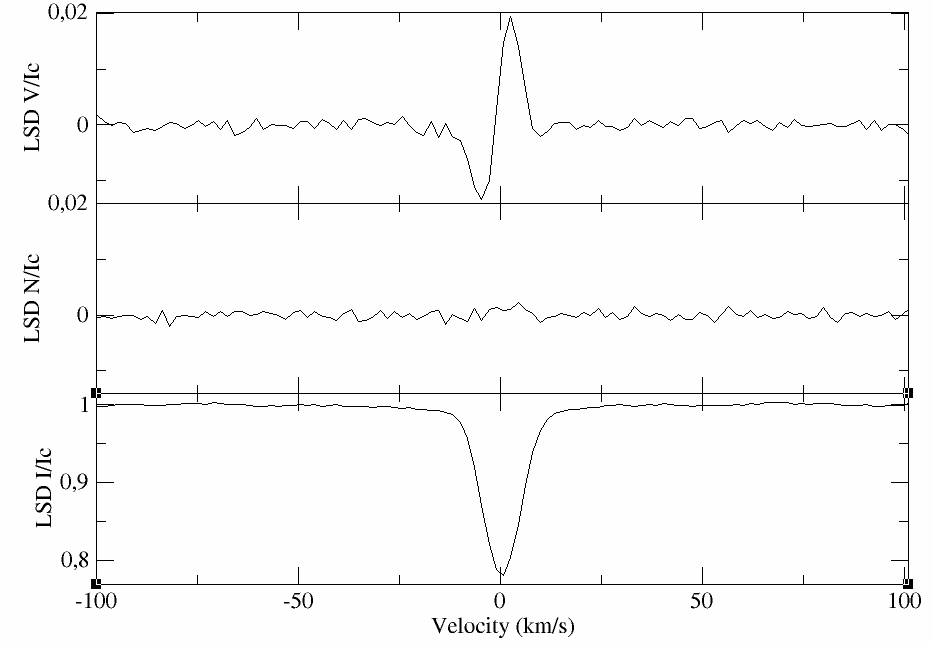}
         \caption{HD\,66533 -- TIC 169971995}
         \label{fig:hd66533_lsd}
     \end{subfigure}
     \vskip\baselineskip
     \begin{subfigure}{0.49\textwidth}
         \centering
         \includegraphics[width=\textwidth]{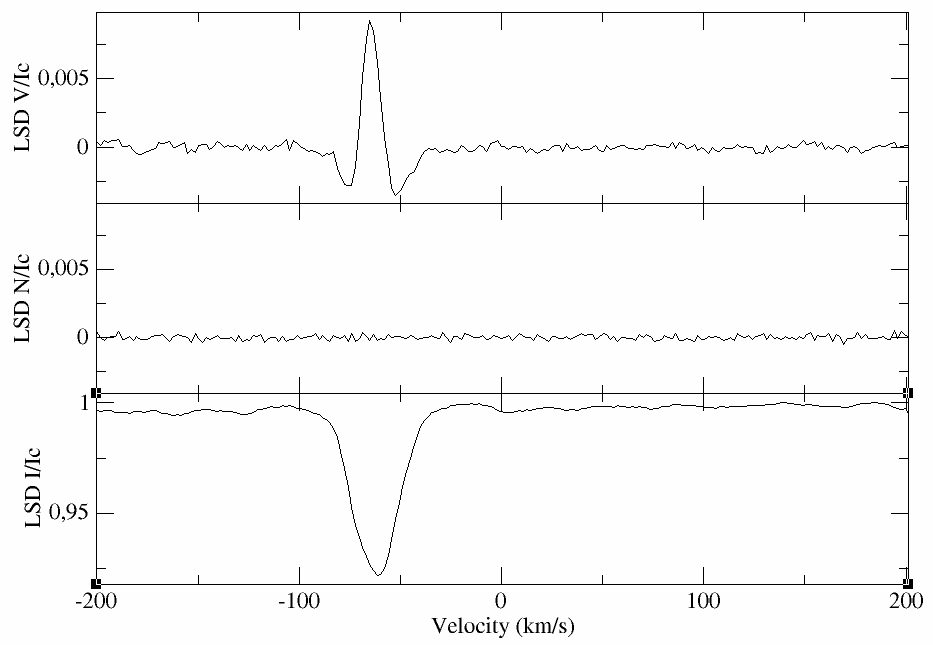}
         \caption{HD\,71047 -- TIC 27256691}
         \label{fig:hd71047_lsd}
     \end{subfigure}
     \hfill
     \begin{subfigure}{0.49\textwidth}
         \centering
         \includegraphics[width=\textwidth]{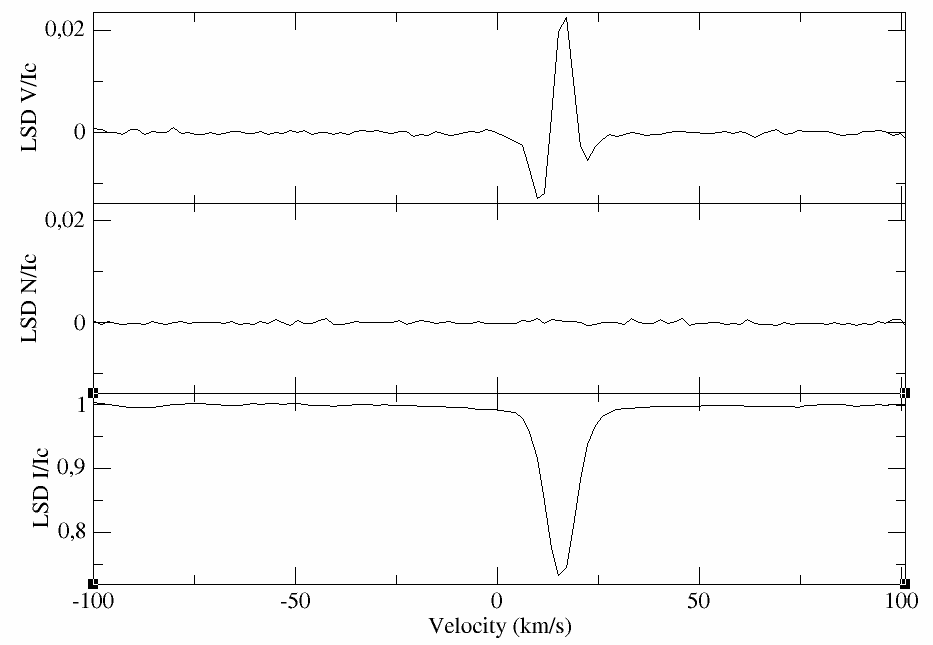}
         \caption{HD\,86170 -- TIC 62815493}
         \label{fig:hd86170_lsd}
     \end{subfigure}
    \caption{LSD Stokes V (upper panel), N (middle panel), and Stokes I (bottom panel) profiles for the various stars in our sample, arranged by stellar identifier (cont.).}
    \label{fig:stokes4}
\end{figure}

\begin{figure}[!ht]
    \centering
    \begin{subfigure}{0.49\textwidth}
         \centering
         \includegraphics[width=\textwidth]{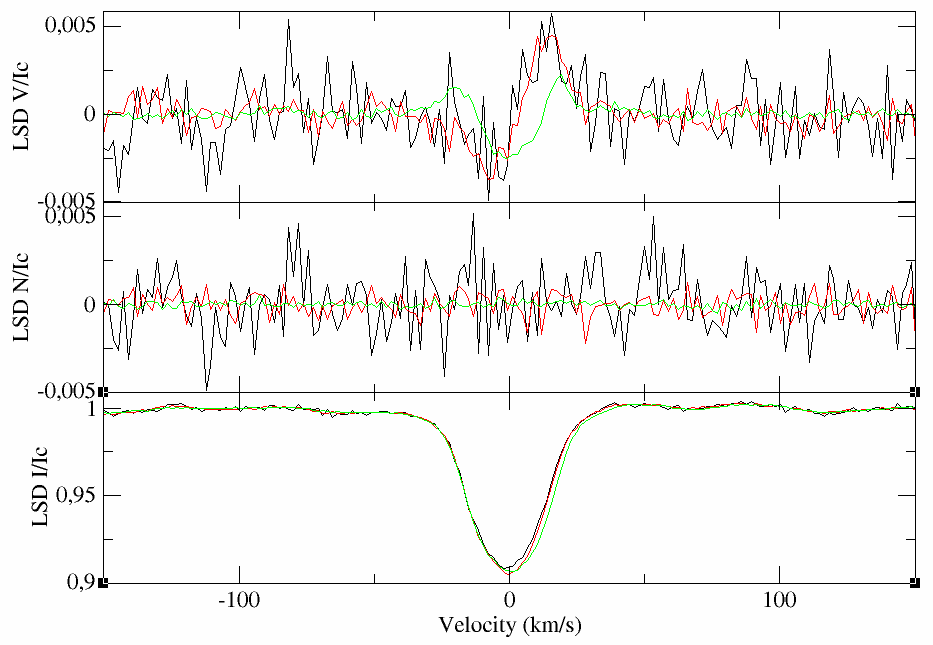}
         \caption{HD\,108662 -- TIC 393808105}
         \label{fig:hd108662_lsd}
    \end{subfigure}
    \hfill
    \begin{subfigure}{0.49\textwidth}
         \centering
         \includegraphics[width=\textwidth]{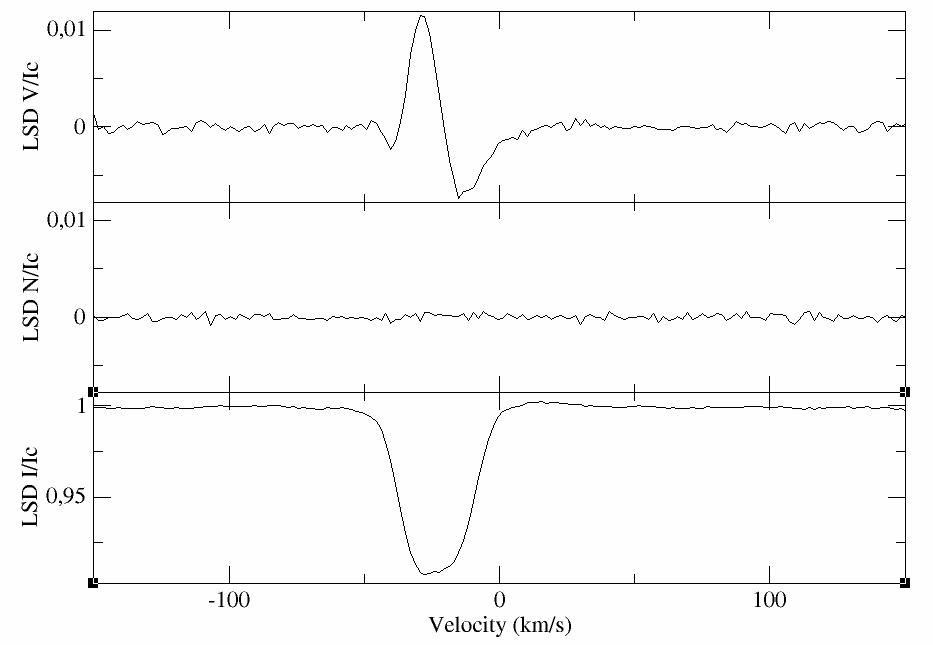}
         \caption{HD\,177128 -- TIC 120495323}
         \label{fig:hd177128_lsd}
     \end{subfigure}
     \vskip\baselineskip
     \begin{subfigure}{0.49\textwidth}
         \centering
         \includegraphics[width=\textwidth]{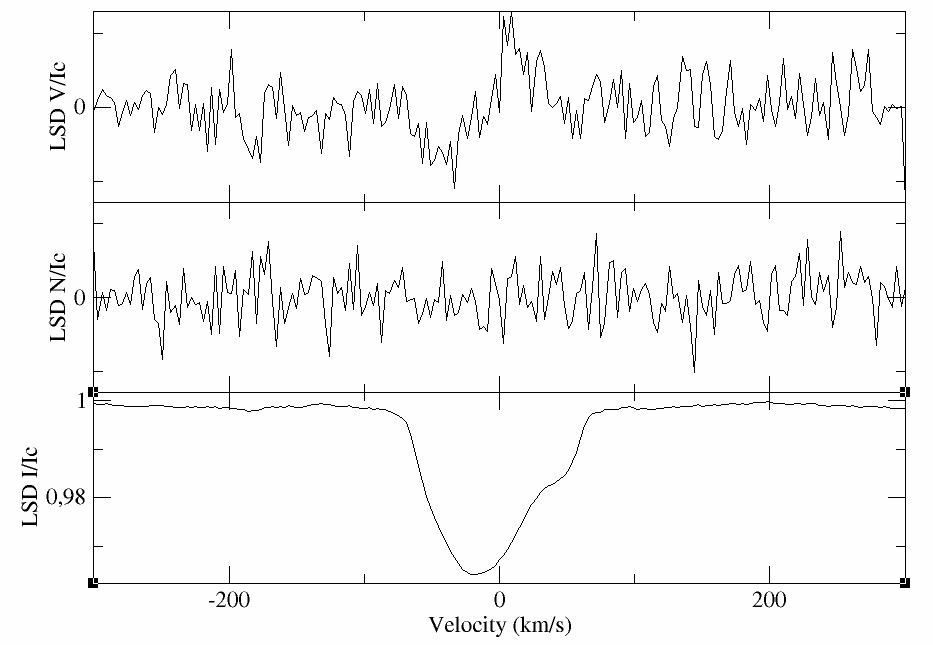}
         \caption{HD\,212714 -- TIC 164282311}
         \label{fig:hd212714_lsd}
     \end{subfigure}
     \hfill
     \begin{subfigure}{0.49\textwidth}
         \centering
         \includegraphics[width=\textwidth]{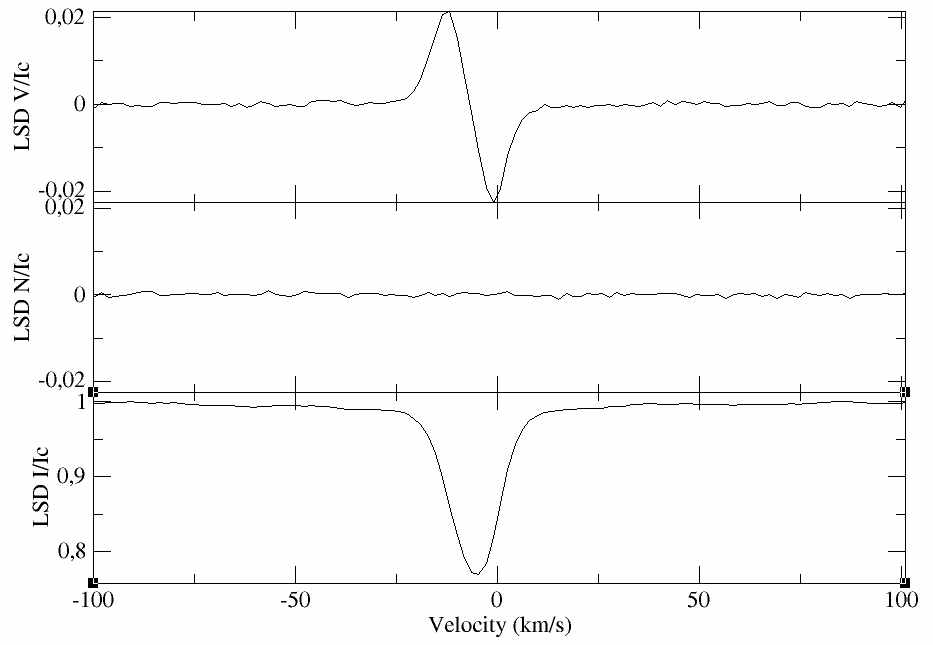}
         \caption{HD\,232285 -- TIC 240808702}
         \label{fig:hd232285_lsd}
     \end{subfigure}
     \vskip\baselineskip
     \begin{subfigure}{0.49\textwidth}
         \centering
         \includegraphics[width=\textwidth]{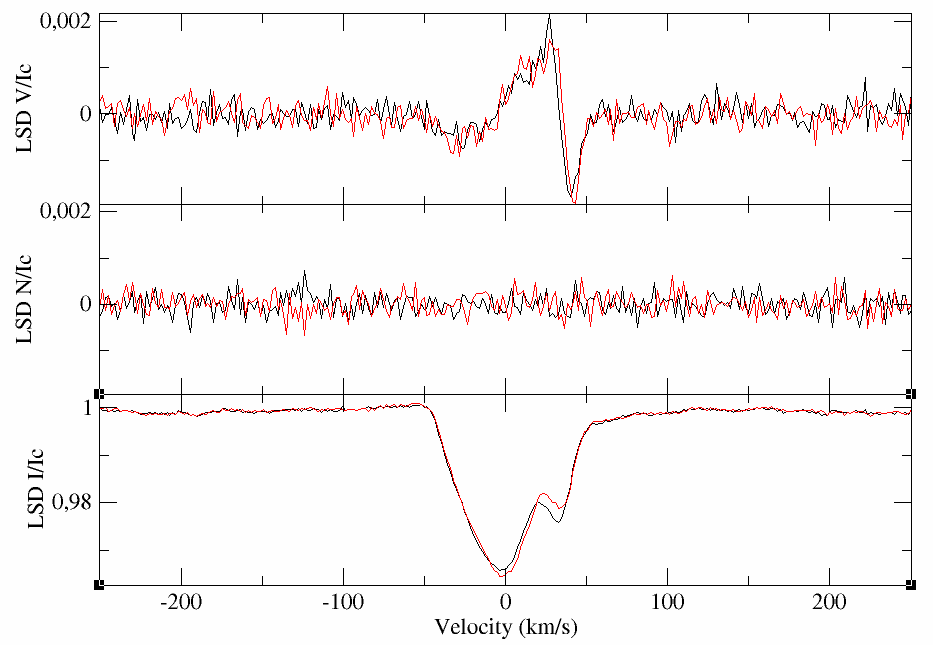}
         \caption{HD\,256582 -- TIC 319616512}
         \label{fig:hd256582_lsd}
     \end{subfigure}
     \hfill
     \begin{subfigure}{0.49\textwidth}
         \centering
         \includegraphics[width=\textwidth]{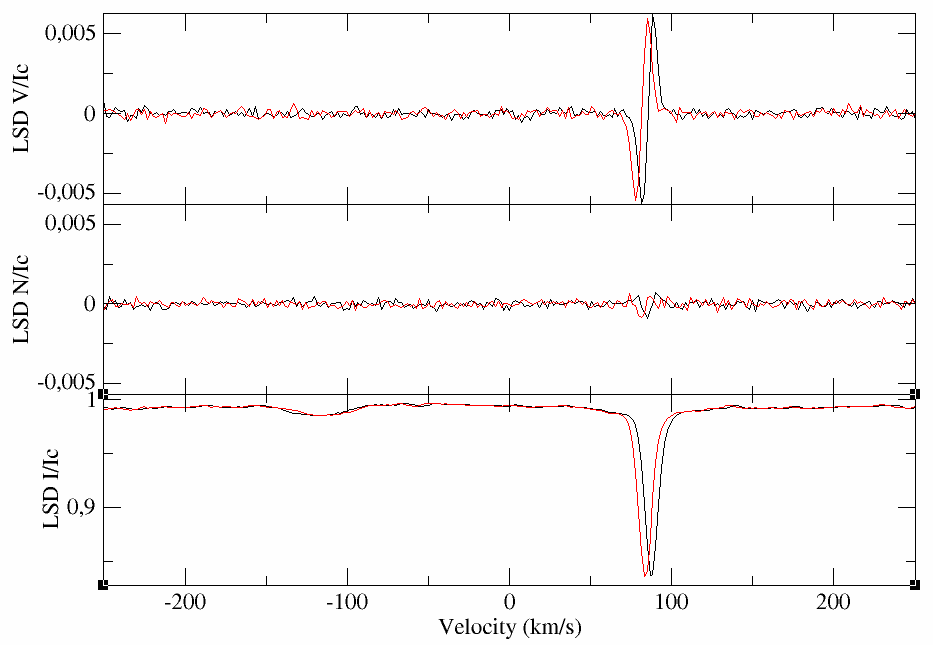}
         \caption{HD\,259273 -- TIC 234878810}
         \label{fig:hd259273_lsd}
     \end{subfigure}
    \caption{LSD Stokes V (upper panel), N (middle panel), and Stokes I (bottom panel) profiles for the various stars in our sample, arranged by stellar identifier (cont.).}
    \label{fig:stokes5}
\end{figure}

\begin{figure}[!ht]
    \centering
    \begin{subfigure}{0.49\textwidth}
         \centering
         \includegraphics[width=\textwidth]{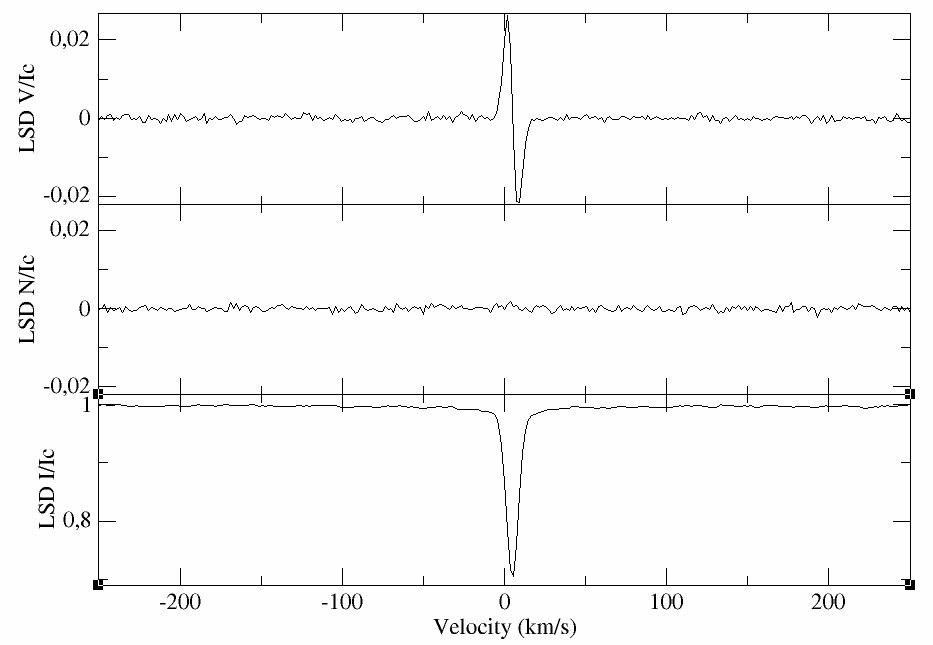}
         \caption{HD\,266267 -- TIC 235391838}
         \label{fig:hd266267_lsd}
    \end{subfigure}
    \hfill
    \begin{subfigure}{0.49\textwidth}
         \centering
         \includegraphics[width=\textwidth]{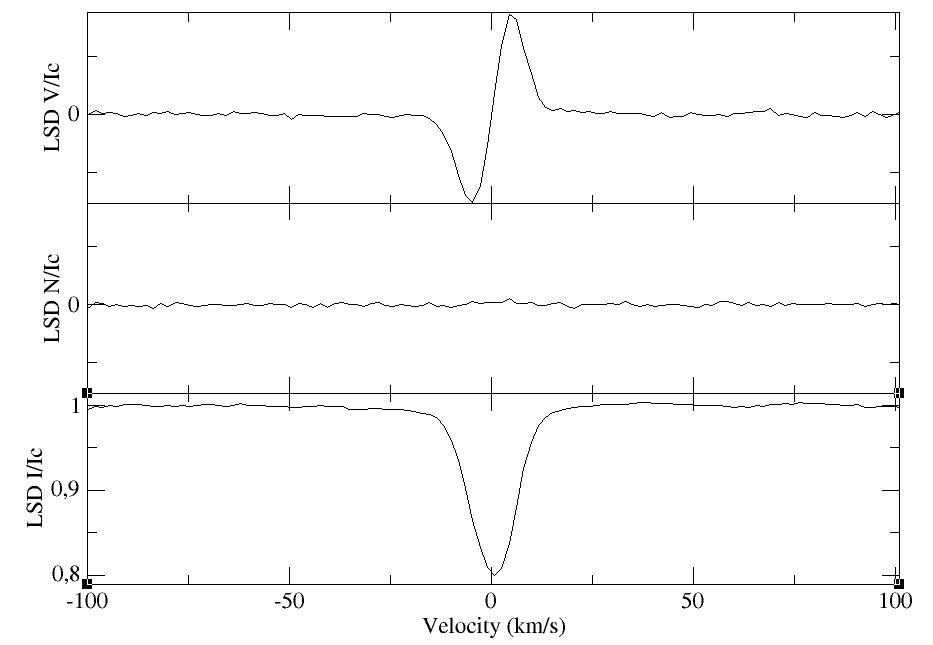}
         \caption{HD\,266311 -- TIC 237662091}
         \label{fig:hd266311_lsd}
     \end{subfigure}
     \vskip\baselineskip
     \begin{subfigure}{0.49\textwidth}
         \centering
         \includegraphics[width=\textwidth]{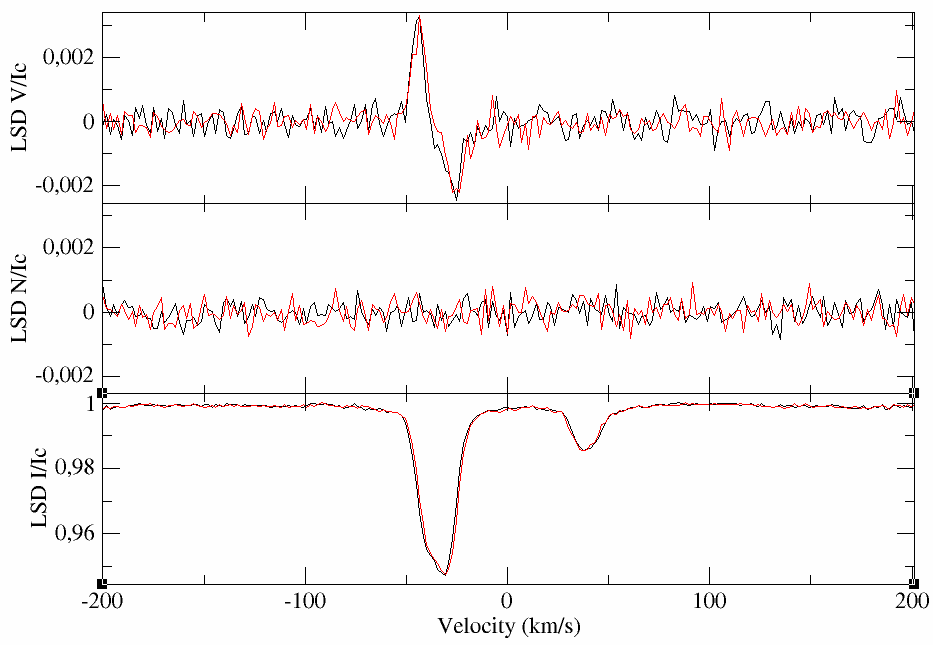}
         \caption{HD\,277595 -- TIC 122563793}
         \label{fig:hd277595_lsd}
     \end{subfigure}
     \hfill
     \begin{subfigure}{0.49\textwidth}
         \centering
         \includegraphics[width=\textwidth]{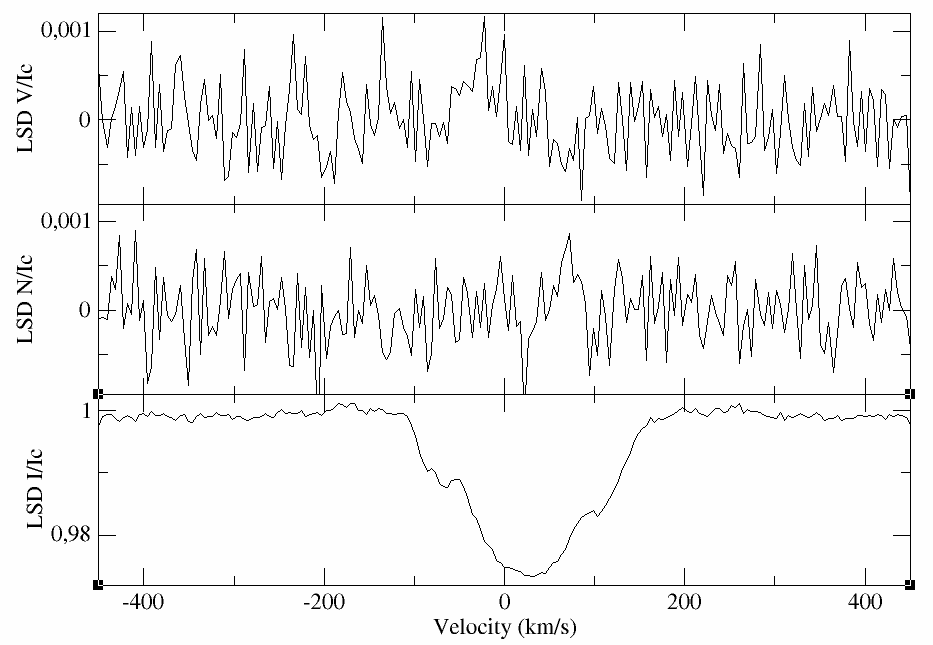}
         \caption{HD\,281193 -- TIC 385555521}
         \label{fig:hd281193_lsd}
     \end{subfigure}
     \vskip\baselineskip
     \begin{subfigure}{0.49\textwidth}
         \centering
         \includegraphics[width=\textwidth]{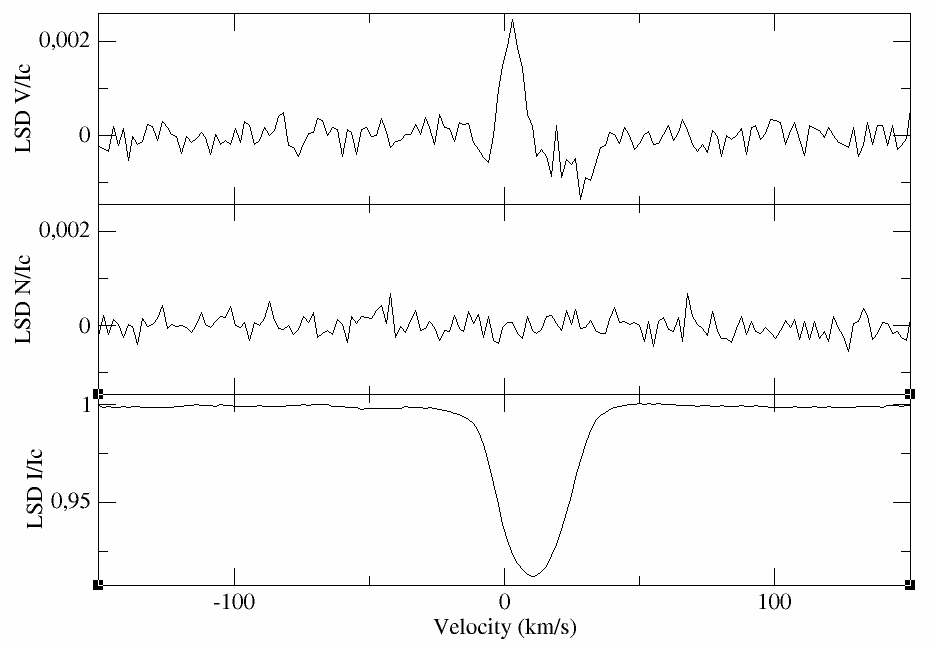}
         \caption{TYC 2873-3205-1 -- TIC 384988765}
         \label{fig:tyc2873_lsd}
     \end{subfigure}
     \hfill
     \begin{subfigure}{0.49\textwidth}
         \centering
         \includegraphics[width=\textwidth]{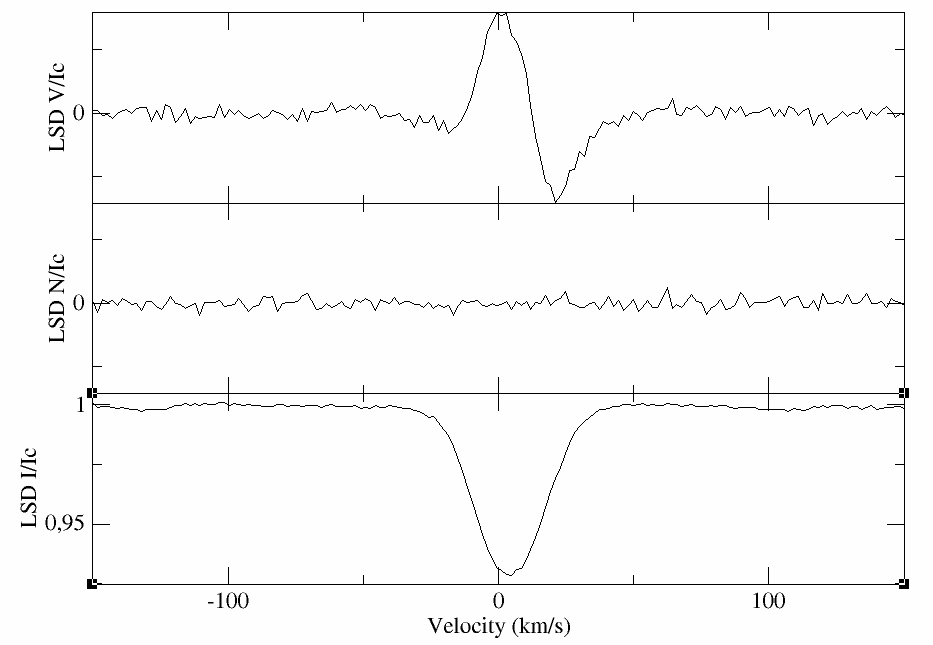}
         \caption{TYC 3316-892-1 -- TIC 458780077}
         \label{fig:tyc3316_lsd}
     \end{subfigure}
    \caption{LSD Stokes V (upper panel), N (middle panel), and Stokes I (bottom panel) profiles for the various stars in our sample, arranged by stellar identifier (cont.).}
    \label{fig:stokes6}
\end{figure}

\begin{figure}[!ht]
    \centering
    \begin{subfigure}{0.49\textwidth}
         \centering
         \includegraphics[width=\textwidth]{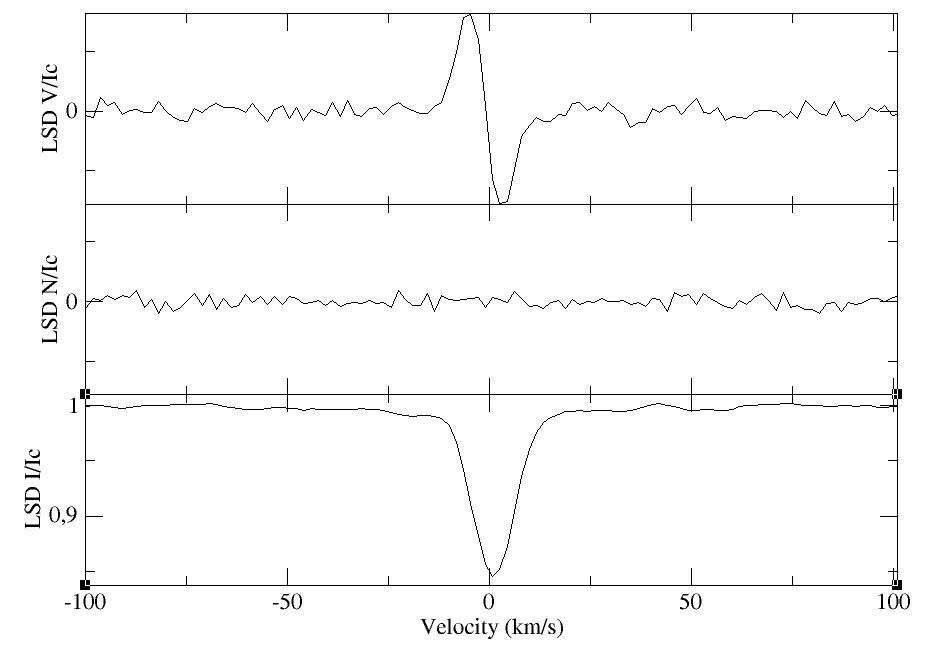}
         \caption{TYC 3319-464-1 -- TIC 117663254}
         \label{fig:tyc3319_lsd}
    \end{subfigure}
    \hfill
    \begin{subfigure}{0.49\textwidth}
         \centering
         \includegraphics[width=\textwidth]{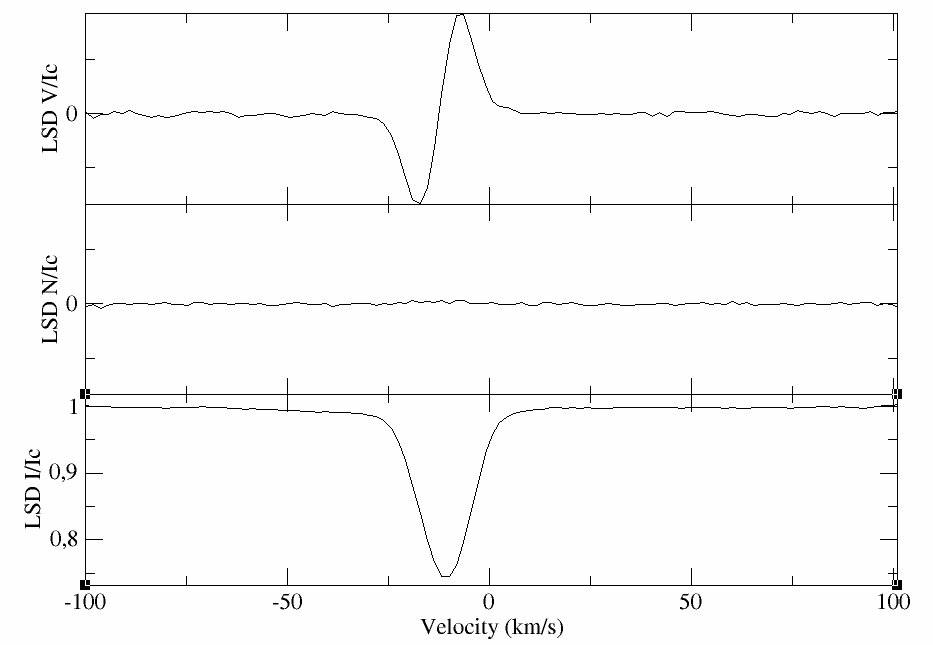}
         \caption{TYC 3733-133-1 -- TIC 252212077}
         \label{fig:tyc3733_lsd}
    \end{subfigure}
    \vskip\baselineskip
    \begin{subfigure}{0.49\textwidth}
         \centering
         \includegraphics[width=\textwidth]{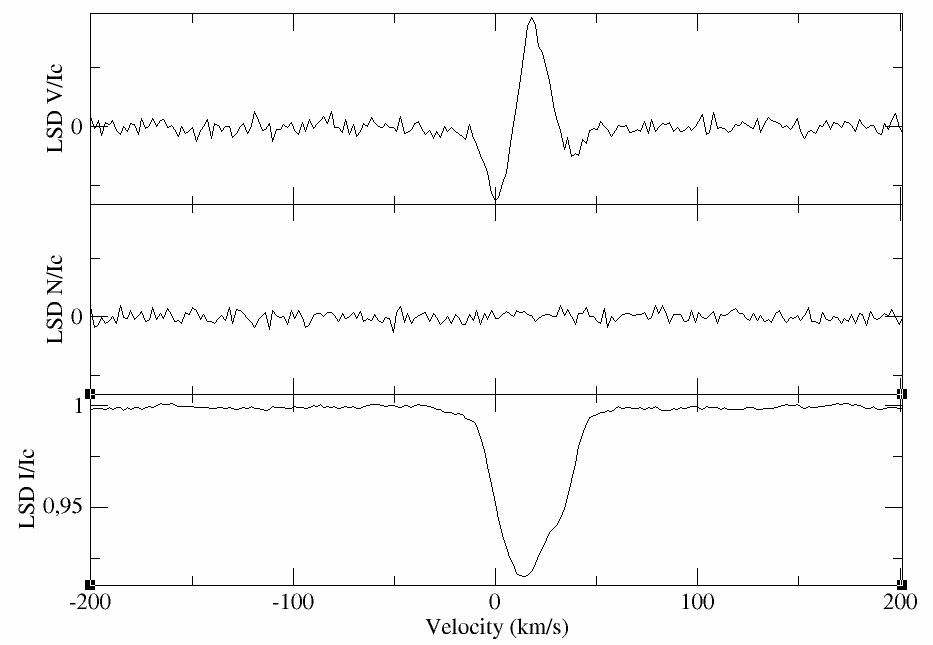}
         \caption{TYC 3749-888-1 -- TIC 321832920}
         \label{fig:tyc3749_lsd}
    \end{subfigure}
    \caption{LSD Stokes V (upper panel), N (middle panel), and Stokes I (bottom panel) profiles for the various stars in our sample, arranged by stellar identifier (cont.).}
    \label{fig:stokes7}
\end{figure}
} 

\newpage
\clearpage
\twocolumn{
\section{\emph{TESS} frequency spectra} \label{apx:TESS}

We revisited the \emph{TESS} data for the four pulsators and one EB of this sample, including more recent data than was presented in \citet{labadie2023} whenever available. For the three $\delta$\,Sct pulsators, it is important to ensure that the stellar oscillation frequencies are below the Nyquist frequency of the \emph{TESS} photometry. Only one of these, HD\,36955 (TIC\,427377135), had 2-minute cadence data (from sector 6), so we largely rely on the full frame images (FFIs). The cadence of the FFIs has increased since the start of the mission, resulting in higher Nyquist frequencies and thus the ability to detect higher frequency signals. In the first two TESS cycles (sectors 1 to 26), the FFI cadence was 30 minutes (Nyquist frequency of 24 d$^{-1}$), in the next two cycles (sectors 27 to 55) the FFI cadence was 10 minutes (Nyquist frequency of 72 d$^{-1}$), and in cycle 56 and later the FFI cadence was increased to 200 seconds (Nyquist frequency of 216 d$^{-1}$)). 

The frequency spectra are plotted in Fig.~\ref{fig:FTs}, computed using Period04 \citep{Lenz2005}. First the rotational frequency and any harmonics were removed, and subsequent peaks were identified through iterative pre-whitening and a simultaneous fit of all significant peaks (above a SNR threshold of 4, using a 2 d$^{-1}$ window centred on each signal).
Inspection of the highest-cadence TESS light curve available for these targets (2-minute cadence for HD\,36955, and 200 second cadence for the rest) did not show any significant peaks beyond what is plotted in Fig.~\ref{fig:FTs}, demonstrating that the 10-minute cadence data do not suffer from super-Nyquist aliasing (although the amplitude of the highest frequency peaks for HD\,36955 may be slightly suppressed in the sector 32 data). The frequency spectrum in Fig.~\ref{fig:FTs} for HD\,49198 supersedes what is shown in \citet{labadie2023} which relied on 30-minute cadence data (and so did suffer from super-Nyquist aliasing). 

\textit{HD\,277595 :} The low-frequency group of signals centred around 1.5 d$^{-1}$ is poorly resolved, and it is unclear as to whether or not there are genuine differences between sectors 19 and 59, or if the perceived differences are due to insufficient frequency resolution. However, the difference in the signals lower than 0.5 d$^{-1}$ do seem genuine. Considering the spectral type (B8VSi), the photometric variations are consistent with SPB pulsation.

\textit{HD\,49198 :} The frequency spectrum is similar when comparing sectors 60 and 73, with some changes in the relative strength of a few signals. The observed behaviour is typical for $\delta$ Scuti pulsators.

\textit{HD\,63843 :} There are significant differences in the relative amplitude of signals that are strong in one or both sectors (e.g. at 19.7 d$^{-1}$ and 16.7 d$^{-1}$). Two higher frequency signals, at 26.5 and 26.8 d$^{-1}$ only appear prominently in sector 61. These signals are typical of $\delta$ Scuti pulsators.

\textit{HD\,36955 :} The frequency spectra are similar in sectors 6 and 32 for the high frequency $\delta$ Scuti pulsation, but with some relative changes in amplitude. The lower frequency signals ($<$ 2 d$^{-1}$) change between sectors and possibly reflect $\gamma$ Dor pulsation, although this is not entirely clear.

\textit{HD\,259273 :}
We re-visited the \emph{TESS} photometry for this eclipsing binary considering the information gleaned from our analysis of the ESPaDOnS data. In \citet{labadie2023}, the rotation period of the magnetic star was presumed to be the same as that of the orbital period (3.409073 d), due to the unambiguous nature of the eclipses and the coherent out-of-eclipse variability consistent with rotation. However, from the very narrow Stokes I line profile for the magnetic component (Fig.~\ref{fig:hd259273_lsd}) we measured $v \sin i = 5.1 \pm 2$ km s$^{-1}$. 
Coarsely estimating the radius of the magnetic star to be between 2 and 3 R$_{\odot}$ (consistent with its late B spectral type), considering the uncertainty on $v \sin i $, and setting $i=90^{\circ}$, the rotation period should be between about 14 and 50 days -- considerably slower than the orbital period and the out-of-eclipse variability. A more careful extraction of the two sectors of available \emph{TESS} data revealed a slow oscillation with a 15.5 day period, as shown by the red curves in Fig.~\ref{fig:EB2}, which may be caused by rotation of the magnetic star. 
Future additional spectropolarimetry can be used to unambiguously determine the rotation period. The secondary is more broad lined with $v \sin i = 5.1 \pm 2$ km s$^{-1}$, where a radius of 1.5 (2.0) R$_{\odot}$ and $i=90^{\circ}$ corresponds to a rotation period of 3.0 (4.0) d, and thus may be tidally locked. 

After subtracting the slow oscillation, the same photometry phased to the orbital period is shown in Fig.~\ref{fig:EB1} along with the RV measurements for both components determined from the Stokes I profile (Fig.~\ref{fig:hd259273_lsd}), and one RV measurement of the primary (magnetic) star from LAMOST DR8 \citep{lamostRVs}. Although the RV measurements are sparse, they suggest that the magnetic star is a genuine member of the eclipsing pair (and not, e.g., a further out tertiary), and that it is the hotter component. 
Since the magnetic star seems to rotate much slower than the orbital period, it is not tidally locked (but the secondary may be). 
A rotating secondary can only explain the variation if it has surface features that are stable for at least as long as the \emph{TESS} observing baseline ($\sim$2 years). Magnetism can cause such spots, but no magnetic field was detected in the secondary (Sec.~\ref{sec:HD259273}).
The out-of-eclipse variation is then perhaps best explained by a reflection effect where the hotter primary irradiates the primary-facing hemisphere of the secondary.
The orbit appears to not be perfectly circular, as the secondary eclipse occurs at 0.510 in phase after the primary eclipse. From the RV semi-amplitudes and the relative eclipse depths, the secondary should be $\sim$0.77 times the mass of the primary, and $\sim$0.9 times the effective temperature of the primary (consistent with an early to mid A star). Further data and analysis are required to properly characterise the stellar and binary properties of this system, but the information presented here provides a useful starting point.




The radial velocity measurements (determined from a Gaussian fit) from the two Stokes I profiles for HD\,259273 are $87.56 \pm 0.05$ km~s$^{-1}$ and $-119.08 \pm 0.5$ km~s$^{-1}$ for the primary and secondary components, respectively, for the observation taken at  MJD = 2459574.846135, and $83.62 \pm 0.05$ km~s$^{-1}$ and $-114.99 \pm 0.5$ km~s$^{-1}$ for the observation at MJD = 2459574.892165.

\begin{figure*}
    \centering
    \includegraphics[width=0.95\textwidth]{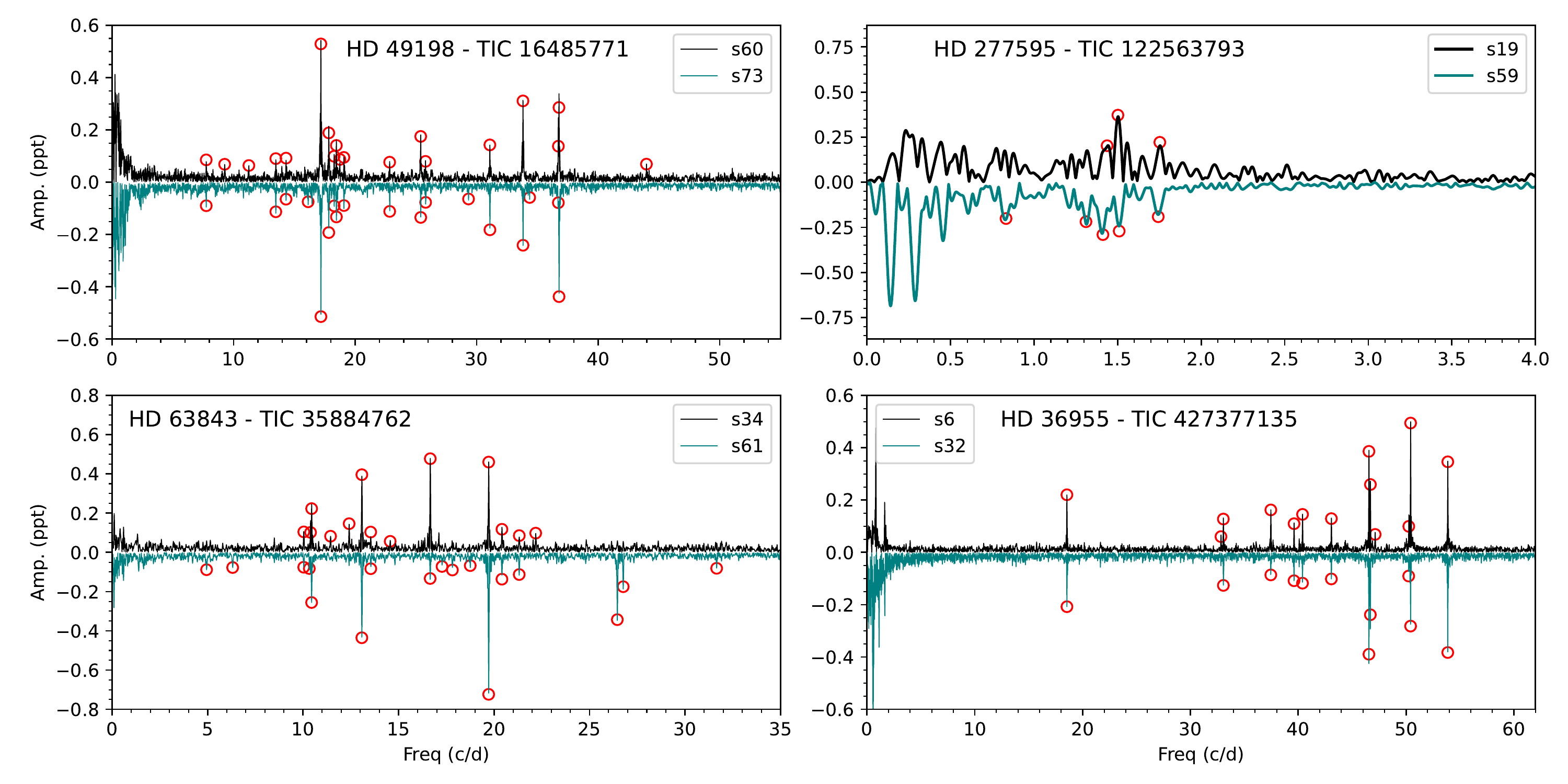}
    \caption{\emph{TESS} frequency spectra for the four pulsators after removing the rotational signal (except for HD\,63843 which does not show detectable rotational modulation in the \emph{TESS} data), computed from two different sectors of data for each star (the curve for the latter sector is inverted). Individual frequencies are marked with red circles. }
    \label{fig:FTs}
\end{figure*}

\begin{figure*}
    \centering
    \includegraphics[width=0.95\textwidth]{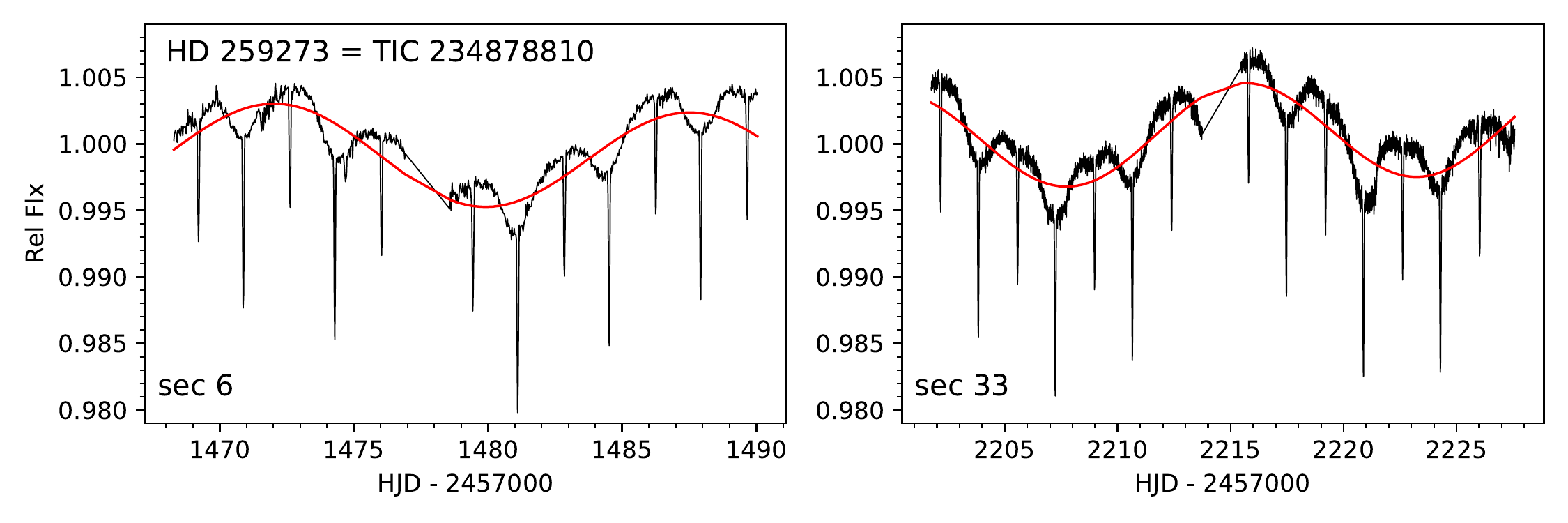}
    \caption{\emph{TESS} light curves of HD\,259273 from sectors 6 and 33. The red curve shows the 15.5 day signal that is potentially associated with rotation of the magnetic star.  }
    \label{fig:EB2}
\end{figure*}

\begin{figure*}
    \centering
    \includegraphics[width=0.95\textwidth]{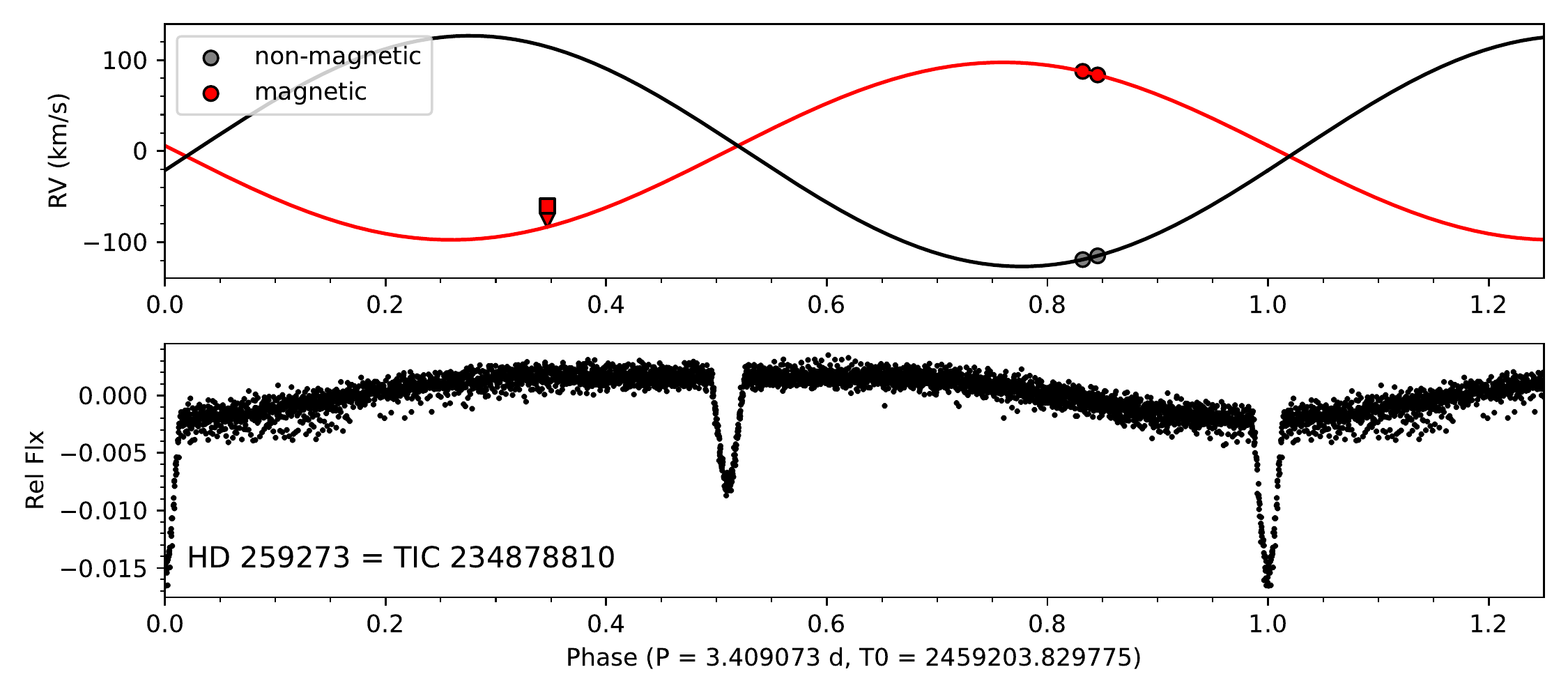}
    \caption{\emph{TESS} light curve of HD\,259273 (bottom) from sectors 6 and 33, phased to the orbital period, after removing the $\sim$15.5 day signal. The upper panel shows the RV measurements from the two ESPaDOnS Stokes I spectra shown in Fig.~\ref{fig:hd259273_lsd} for both components (red and black circles for the magnetic and non-magnetic stars respectively). The RV measurements from LAMOST are indicated by a red square (for the red arm of the spectrograph) and triangle (for the blue arm). A sinusoid is fit to the ESPaDOnS RVs to give a qualitative impression of the expected RV curves, but is not quantitatively reliable with just two closely-spaced measurements. }
    \label{fig:EB1}
\end{figure*}

}

\end{appendix}

\end{document}